\documentclass[showpacs,preprintnumbers,amsmath,amssymb,aps,nofootinbib]{revtex4}

\usepackage{graphicx}
\usepackage{dcolumn}
\usepackage{bm}
\usepackage{multirow}
\usepackage{subfigure}

\def\LL{\left\langle}	
\def\RR{\right\rangle}	

\newcommand{\BE}{\begin{equation}}
\def\EE{\end{equation}}
\def\BEA{\begin{eqnarray}}
\def\EEA{\end{eqnarray}}

\newcommand{\ie}{{\em i.e.\ }}

\newcommand{\Tr}{{\rm tr}}

\newcommand{\meff}{m_\mathrm{eff}}
\newcommand{\tmin}{t_\mathrm{min}}
\newcommand{\tmax}{t_\mathrm{max}}
\newcommand{\chisq}{{\chi^{2}}}
\newcommand{\chisqpdof}{\chisq/\mathrm{d.o.f.}}
\newcommand{\modp}{\left|\vec{p}\right|}
\newcommand{\uubar}{u\overline{u}}
\newcommand{\ddbar}{d\overline{d}}
\newcommand{\ssbar}{s\overline{s}}
\newcommand{\op}{\mathcal{O}}
\newcommand{\Pop}{\mathcal{P}}
\newcommand{\Hop}{\mathcal{H}}
\newcommand{\Rep}{\mathcal{R}}
\newcommand{\App}{A_1^{++}}
\newcommand{\Amp}{A_1^{-+}}
\newcommand{\Epp}{E_1^{++}}
\newcommand{\ord}[1]{\mathcal{O}(#1)}

\begin{document}
\preprint{LTH 876}

\title{Glueball mass measurements from improved staggered fermion simulations}

\author{Christopher M. Richards, Alan C. Irving} 
\affiliation{Theoretical Physics Division, Department of Mathematical Sciences,
        University of Liverpool, Liverpool L69-7ZL, United Kingdom}

\author{Eric B. Gregory} 
\affiliation{Department of Physics, University of Cyprus, P.O. Box 20357
1678 Nicosia, Cyprus}

\author{Craig McNeile} 
\affiliation{Bergische Universit\"at Wuppertal, Gaussstr.\,20, D-42119 Wuppertal, Germany}

\bigskip
\author{\em (UKQCD collaboration)}

\bigskip
\date{May 14, 2010}
    
\begin{abstract}
We present the first 2+1 flavour spectrum measurements of glueball states using 
high statistics simulations with improved staggered fermions.
We find a spectrum consistent with quenched measurements of scalar, pseudoscalar and
tensor glueball states. 
The measurements were made using $5000$ configurations
at a lattice spacing of $0.123$ fm and pion mass of $280$ MeV 
and $3000$ configurations at $0.092$ fm 
with a pion mass of $360$ MeV. 
We see some evidence of coupling to $2\pi$ states. 
We compare our results with the experimental glueball candidate
spectrum as well as quenched glueball estimates. 

\end{abstract}
   

\maketitle

\section{INTRODUCTION} \label{se:INTRODUCTION}
During the last decade, a major first-principles calculation of the
hadron spectrum has been carried out based on the improved staggered
fermion discretisation formalism~\cite{Orginos:1998ue,Aubin:2004wf}.
Comparisons with experiment, have in
most cases shown impressive agreement, typically within a few percent
accuracy. See for example \cite{Davies:2003ik} and references therein
and a recent review article by the MILC collaboration~\cite{Bazavov:2009bb}.
Noticeably absent from this impressive list are some singlet quantities such
as those contributing to glueball states and to the $\eta/\eta'$
system. Since the latter states are stable under the strong
interaction, they may be classified as \lq gold-plated\rq{} 
quantities~\cite{Davies:2003ik} and so deserve urgent consideration
as a test of this otherwise highly successful formalism. This is made
all the more pressing in view of the continuing controversy over the
validity of the so-called \lq fourth-root trick\rq{} for dealing with
the spurious taste multiplicity. See \cite{Sharpe:2006re} for a
comprehensive review of this. A major difficulty in simulating
singlet quantities is the need to make high quality stochastic
estimates of disconnected correlators which are typically very noisy 
observables.
In dynamical simulations the problem is particularly severe due to
the high simulation costs. An analysis of these measurement problems
and initial results using MILC configurations~\cite{Bernard:2001av}
has recently been presented~\cite{Gregory:2007ev}. We have now
completed significantly higher statistics simulations at two lattice
spacings primarily in order to study the $\eta/\eta'$. (These results will be
presented elsewhere.) As a by-product, we have also made measurements of
scalar meson operators and glueball operators. The latter form the
basis of the present paper.

In the quenched approximation, comprehensive studies of the glueball 
spectrum have been available for some
time~\cite{Bali:1993fb,Morningstar:1999rf,Chen:2005mg}.
In comparison, studies with dynamical quarks are still at a 
preliminary stage~\cite{Hart:2001fp,Bali:2000vr,Hart:2006ps,Gregory:2008mn}.
This is mainly because glue correlators are noisy and the masses
relatively high so that signal to noise is hard to control without
large statistics. The quenched studies typically used thousands of
configurations rather than the \lq few hundred\rq{} configurations
usually available at a given lattice spacing and quark mass.
We therefore took the opportunity to make use of several thousand
configurations generated as part of the main singlet simulation
program. Preliminary results were presented in~\cite{Gregory:2008mn}.

Crede and Meyer~\cite{Crede:2008vw} 
and Klempt and Zaitsev~\cite{Klempt:2007cp}
have recently reviewed the past and future experimental
searches for glueballs.
It is particularly timely to study the glueball spectrum
using unquenched QCD, because there are new experiments
starting, or starting soon, that are looking for glueballs.
The BES-III experiment~\cite{Bian:2009zzd} has begun running
and can search for glueballs via radiative $J/\psi$ decays.
It will study light mesons with $0^{-+}$, $2^{++}$, and
$0^{++}$ quantum numbers, where glueball degrees
of freedom could contribute.

It will be particularly interesting to see if the existence of 
$f_J(2220)$ as a state is confirmed by BES-III~\cite{Bian:2009zzd}. 
The $f_J(2220)$ is a candidate $2^{++}$ state that some 
speculate is a glueball state because its mass is close to
the mass of the $2^{++}$ glueball, $(2390(30)(120)$ MeV,
in the quenched estimates of Chen et al.~\cite{Chen:2005mg}. 
The $120$ MeV error is from different ways of
setting the lattice spacing in quenched QCD, parametrised
by uncertainty in $r_0$. Recently the HPQCD determined
$r_0$ to $0.8\%$ accuracy from 
unquenched lattice QCD~\cite{Davies:2009tsa}. 
Using HPQCD's central value of $1/r_0 = 423$ MeV, changes
the mass of the $2^{++}$ glueball to $2470(30)$ MeV. 
This shift is within the errors of the quenched 
estimate~\cite{Chen:2005mg}, but illustrates
the need for unquenched calculations.

The PANDA experiment at FAIR will look for glueballs
in the range $2.2$ to $5.5$ GeV~\cite{Lutz:2009ff,Messchendorp:2010vk}
after 2015.


The general plan of the paper is as follows.
In section~\ref{se:sim_meas} we describe
the simulations, the ensembles and the methods used for
extracting spectrum information.
The next section~\ref{se:results} contains the results from
these calculations. 
The paper ends with a discussion of these results and some conclusions
in section~\ref{se:conclusions}.

\section{Simulation and measurements}
\label{se:sim_meas}

\subsection{Configuration ensembles}
Available resources allowed us to produce an initial two ensembles
corresponding to existing MILC ensembles, but with significantly greater
statistics and making use of an exact algorithm, 
the RHMC~\cite{Clark:2003na,Clark:2006wp} 
rather than the R algorithm. 
The simulation parameters are summarised in 
Table~\ref{tab:ensparams}.
\begin{table}[!ht]
\begin{center}
\begin{tabular}{|l|lll|lll|rr|}
\hline
ensemble & $N_f$ & $\beta$ & $L^3\times T$ & $am_{l/s}$ & ${r_0}/{a}$ & $a$ [fm]
& $N_\mathrm{cfg}$ & $N_\mathrm{traj}$ \\
\hline
coarse & 2+1 & 6.75  & $24^3\times 64$ & $0.006/0.03$ & $3.8122(74)$ &
$0.12250(24)$ & $5237$ & $31422$ \\
fine & 2+1 & 7.095 & $32^3\times 64$ & $0.00775/0.031$ & $5.059(10)$ &
$0.09230(19)$ & $2867$ & \
$17202$ \\
\hline
\end{tabular}
\caption{Ensembles generated for flavour singlet studies.}
\label{tab:ensparams}
\end{center}
\end{table}

We used an improved version of the RHMC algorithm~\cite{Clark:2006wp}
which made use of such things as the \lq$n^{\rm th}$-root trick\rq{},
higher order integrators and different types of 
mass-preconditioning. Additional parameters to be tuned
included gauge/fermion step sizes and the conjugate gradient tolerance
used in different parts of the force calculation and acceptance tests.
Details of the tuning procedures and other properties of the 
simulations are given in~\cite{CMRT}.

The lattice spacing was determined via the static potential
using standard methods~\cite{Allton:2001sk}. 
When required, the value of $r_0$ used to convert to physical scales
was $0.467$ fm.~\footnote{
A recent accurate determination by the HPQCD
collaboration using MILC ensembles gave
$r_0=0.4661(38)$~\cite{Davies:2009tsa}}

The \lq coarse\rq{} ensemble summarised in
Table~\ref{tab:ensparams} represents some $8$ times the statistics of
the corresponding $20^3\times 64$ MILC configuration 
studied in~\cite{Gregory:2007ev}
while the fine ensemble represents an increase of around $5$ on a typical
MILC ensemble. We have studied the autocorrelation time for simple
operators. For technical reasons the first simulations (coarse
ensemble) which were conducted on the UKQCD QCDOC machine~\cite{Boyle:2005gf}
were interrupted at various times resulting in non-contiguous
RHMC trajectories. The ensemble was produced in two separate streams.
The fine lattice simulations however resulted in an
uninterrupted Markov chain, so allowing investigation of
some autocorrelations. Table~\ref{tab:ACtime} shows the integrated
autocorrelation times of  some relevant observables which can be
meaningfully defined on a single configuration.
\begin{table}[!ht]
\begin{center}
\begin{tabular}{|l|l|l|l|}
\hline
              &$LL$               &$LF$         &$FF$\\
\hline
$\meff^{\pi}$  &$2.28(33)$         &$2.29(32)$   &$1.99(21)$\\
\hline
\hline
              &$i=0$              &$i=1$        &$i=2$\\
\hline
$\Pop^{\App}_i(0)$         &$0.65(6)$          &$0.89(21)$   &$1.12(9)$\\
\hline
$\sum_t\Pop^{\App}_i(t)$   &$1.13(15)$         &$1.61(22)$   &$2.59(26)$\\
\hline
\end{tabular}
\caption{Integrated autocorrelation times (in units of $6$
  trajectories) for the effective mass of the pion (averaged over
  euclidean times $8-10$) and for the plaquette based glue
  operators-defined below for a fixed time plane ($t=0$) and also
  averaged over all time planes.}
\label{tab:ACtime}
\end{center}
\end{table}
The effective masses (defined below) are those using both local
($L$) and fuzzed($F$) sources and sinks for the
$\gamma_5\times\gamma_5$ pion operator. The glue operators are
defined in the next section. The labels $i=0,1,2$ refer to the number
of Teper blocking levels as described below. As expected, 
the pion shows a longer autocorrelation than a single time plane
operator. We have not detected any evidence of autocorrelation in
the glueball correlators and corresponding effective masses 
which, as discussed below, are very noisy. 
We have checked, by binning in selected cases, 
that the statistical error estimates made using configurations
separated by 6 trajectories are reasonable. 

The coarse ensemble has been successfully incorporated by the MILC 
collaboration into an
analysis of the strange quark content of the nucleon~\cite{Toussaint:2009pz}.

\subsection{Glueball measurement methods}
\label{sse:meas}
We used standard glue operators $\Pop^{\App}$ built from spatial plaquettes for the $O_h$ irrep $A_1^{++}$ 
(coupling to $0^{++}$ in the continuum). 
Prior to measurement we use APE
smearing~\cite{Albanese:1987ds} (twice with smearing parameter
$c=2.5$) 
and then Teper
blocking~\cite{Teper:1987wt} performed $n$ times where $n=0,1,2,\dots N-1$. 
Using all $N$ blocking levels provides an $N$-dimensional 
basis for the variational
measurement techniques as described below.
The APE smearing smooths out some of the ultra-violet noise while the 
different Teper blocking levels provide a basis of operators with a
range of physical extents allowing different couplings to the ground
state and excited states in a given channel. This is important when
using variationally motivated techniques as outlined below.
We also use these plaquette operators $\Pop^{\Rep}$
for the $E_1^{++}$ and 
$T_2^{++}$ irreps ($\Rep$) which couple to $2^{++}$ states in the continuum.
For example, for $E_1^{++}$ we use both the operators in (\ref{eq:tensorgbop})
(standard) and (\ref{eq:tensorgbop2}) (alternative).
\begin{equation}
\label{eq:tensorgbop}
\mathcal{P}^{E_{1}^{++}}_i(t)=\Tr\sum_{\vec{x}}\left(P^i_{4,xy}(\vec{x},t)
-P^i_{4,yz}(\vec{x},t)\right)
\end{equation}
and
\begin{equation}
\label{eq:tensorgbop2}
\mathcal{P}^{\prime
  E_1^{++}}_i(t)=\Tr\sum_{\vec{x}}\left(P^i_{4,xy}(\vec{x},t)+
 P^i_{4,yz}-2P^i_{4,zx}(\vec{x},t)\right)\, .
\end{equation}

\begin{figure}[!htbp]
\begin{center}
\includegraphics[width=0.35\textwidth,clip]{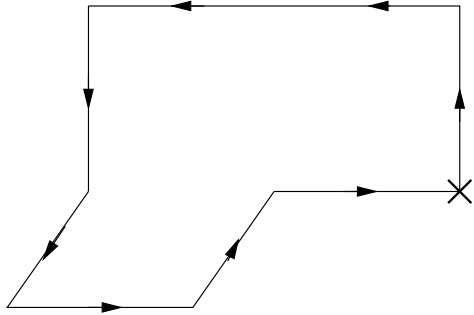}
\caption[Schematic of Handed Wilson Loop]{Schematic of the (left) hand
operator $\Hop_0$ which can be used to study the pseudoscalar
glueball.}
\label{fig:hand_op}
\end{center}
\end{figure}

For pseudoscalar operators ($A_1^{-+}$) we require something
non-planar with some \lq handedness\rq{}~\cite{Berg:1982kp}.  
In practice, see (\ref{eq:psgbop}), we use left ($L$) and 
right ($R$) handed versions of the
loop shown in Fig~\ref{fig:hand_op}.
\begin{equation}
\label{eq:psgbop}
\begin{split}
\mathcal{H}^{A_{1}^{-+}}_i(t)&=\Tr\sum_{\vec{x}}\Big(\left(H^i_{8,L,xy}(t)+
H^i_{8,L,yz}(t)+H^i_{8,L,zx}(t)\right)\\
    &\quad\quad
    -\left(H^i_{8,R,xy}(t)+H^i_{8,R,yz}(t)+H^i_{8,R,zx}(t)\right)\Big)\, .
\end{split}
\end{equation}
We refer to operators $\Hop^{\Rep}$ based on these as \lq hand\rq{}
operators in what follows.

We calculate correlators (vacuum-subtracted where necessary) between
operators with definite momentum on time planes separated by euclidean
time $t$
\BE
C_{ij}^{\Rep}(t)=\LL \op^{\Rep}_i(0)\op^{\Rep}_j(t)\RR\, .
\label{eq:Cij}
\EE
Here $\Rep$ refers to the $O_h$ irrep under consideration and $i,j$ refer
to the operator basis being used e.g. different Teper blocking levels.
These matrices of correlators form the basis of subsequent analyses as
described below.

\begin{enumerate}
\item{\em Effective masses.} These are defined generically by
\BE
\label{eq:meff}
a\meff=\ln\frac{C(t)}{C(t+1)}\, .
\EE
where $C(t)$ is $C_{ij}^{\Rep}(t)$ for some
choice of $\Rep$ and $i=j$. 
Typically, a plateau in $a\meff(t)$ is observed at lower values of 
$t$ when $i$ corresponds to higher levels of blocking. 
Weighted averages over the plateau region can then be used to 
estimate the ground state mass in channel $\Rep$. When
using an improved gauge action containing links over more than one
time-plane, as here, one should be alert to possible positivity violations at
small euclidean time~\cite{Luscher:1984is}. However, we have not encountered
any unusual behaviour in this regard.

\item{\em Variational methods.} In the basic variational 
method~\cite{Michael:1982gb,Blossier:2009kd,Necco:2003vh} 
one finds eigenvalues of the $N\times N$ 
matrix
\BE
M^{\Rep}_{ij}(t,t_0)\equiv \sum_{k=1}^N[C^{\Rep}(t_0)]_{ik}^{-1}C^{\Rep}_{kj}(t)
\EE
where $t_0$ is some initial euclidean time at which the 
correlator matrix is sufficiently well-determined to be invertible.
It is straightforward to show (see for example 
the factorising fit parametrisation
given below) that the eigenvalues $\lambda^\alpha(t_0,t)$ of 
$M^{\Rep}(t_0,t)$ are 
related to the transfer matrix and
\BE
\lambda^\alpha(t_0,t)=e^{-m_\alpha(t-t_0)}
\qquad (\alpha=0,1,\dots N-1) \,.
\label{eq:var_mass}
\EE
One can then either 
\begin{enumerate}
\item use directly the masses $am^{\alpha}$ 
obtained from (\ref{eq:var_mass}) or
\item 
form an effective mass, as in (\ref{eq:meff}) above, from the 
ground state ($\alpha=0$) projection of $C(t)$\, . 
\end{enumerate} 
For convenience, we refer to masses estimated via 
(a) as \lq eigenvalue masses\rq{} and those estimated via (b) as
\lq variational effective masses\rq{}.

\item{\em Factorising fits.} Using the usual intermediate state arguments,
the correlator matrix (\ref{eq:Cij}) can be
expressed as an infinite sum of exponential contributions and fitted
in truncated form so as to extract mass estimates:
\BE
C_{ij}(t) = \sum_{\alpha=0}^{M-1} c^{\alpha}_i c^{\alpha}_j
 e^{-m_{\alpha}t}\, . 
\EE
Here we have suppressed the irrep label $\Rep$ and restricted the sum to
the $M$ lowest-lying states. 
Note that by choosing $M=N$, the dimension of the operator basis, one
can simply recover the variational formula given above. 
The factorised coefficient $c^{\alpha}_i$ gives the overlap of state 
$\alpha$ with operator $i$.
\end{enumerate} 

Note that in this study we have included only glue-based operator
correlations in our factorising fits and corresponding variational
analyses. In section~\ref{se:results} we will comment further on the
prospects for studies of two meson operators and decay studies in
general.

\section{Results} \label{se:results}
\subsection{Results for the scalar glueball -- coarse ensemble} 
\label{sse:coarse}
\begin{figure}[ht]
\begin{center}
\includegraphics[width=0.9\textwidth,clip]
{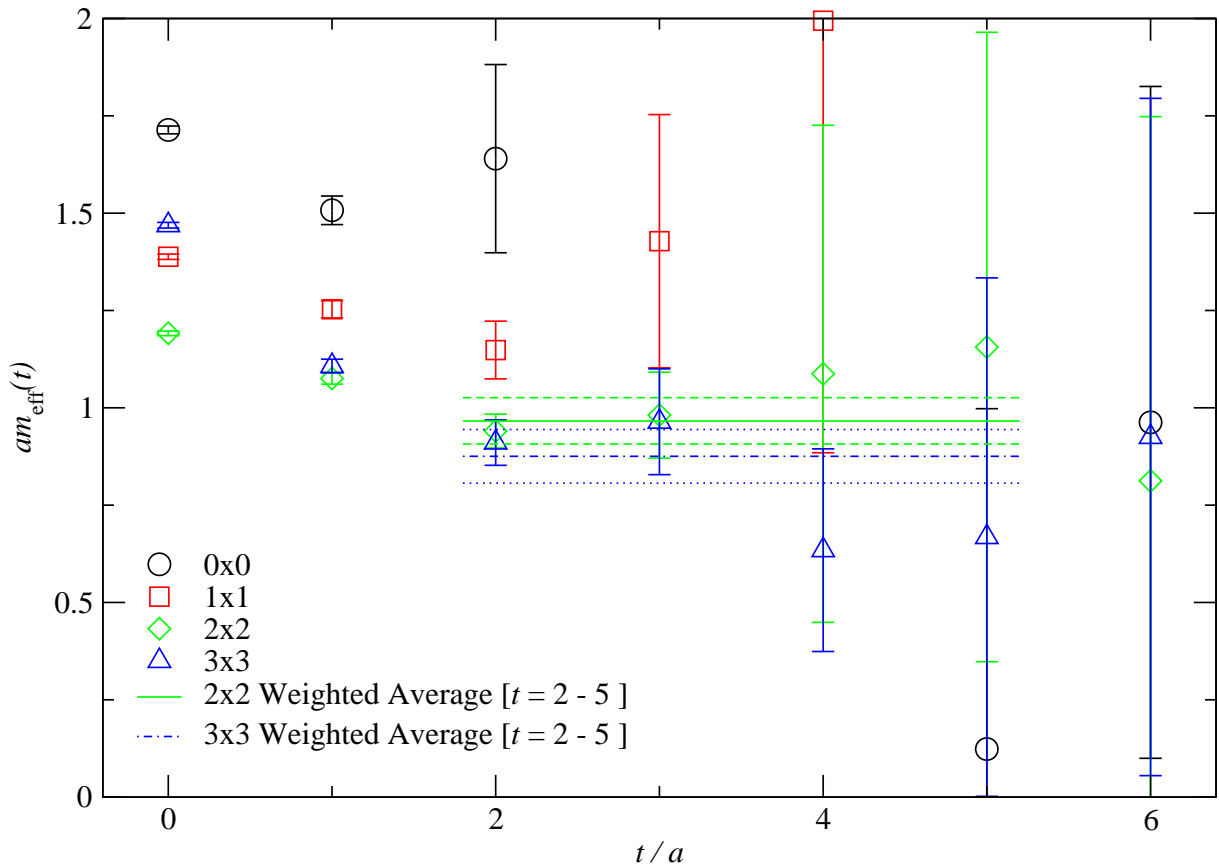}
\caption[Effective mass plot: Scalar Glueball (Coarse)]{The effective
 masses for the diagonal entries $(0,0)$, $(1,1)$ etc. of a 
$4\times 4$ matrix of correlators 
formed using the
$|p|=0$ standard scalar operators  with blocking levels $0$ to $3$ 
measured on the coarse ensemble. Weighted averages for 
blocking levels 2 and 3 are shown, computed from $t=2 - 5$.}
\label{fig:coarse_meff}
\end{center}
\end{figure}

We first give some sample results of the above methods applied to 
the coarse lattices.
Fig.~\ref{fig:coarse_meff} shows the basic effective masses of the
momentum $0$ scalar state for each of the 
blocking levels $0$ to $3$. 
Despite the significantly larger statistics, compared with previous
dynamical simulations, it is clear that the ratio of 
signal to noise is still a significant problem.
Weighted averages of the \lq plateau\rq{} values are shown for blocking
levels $2$ and $3$. Here the weighting is inversely proportional to 
the statistical error. As with all other quantities in this study,
overall statistical errors were estimated via bootstrap. 

\begin{figure}[!htb]
\centering
\includegraphics[width=0.8\textwidth,clip]
{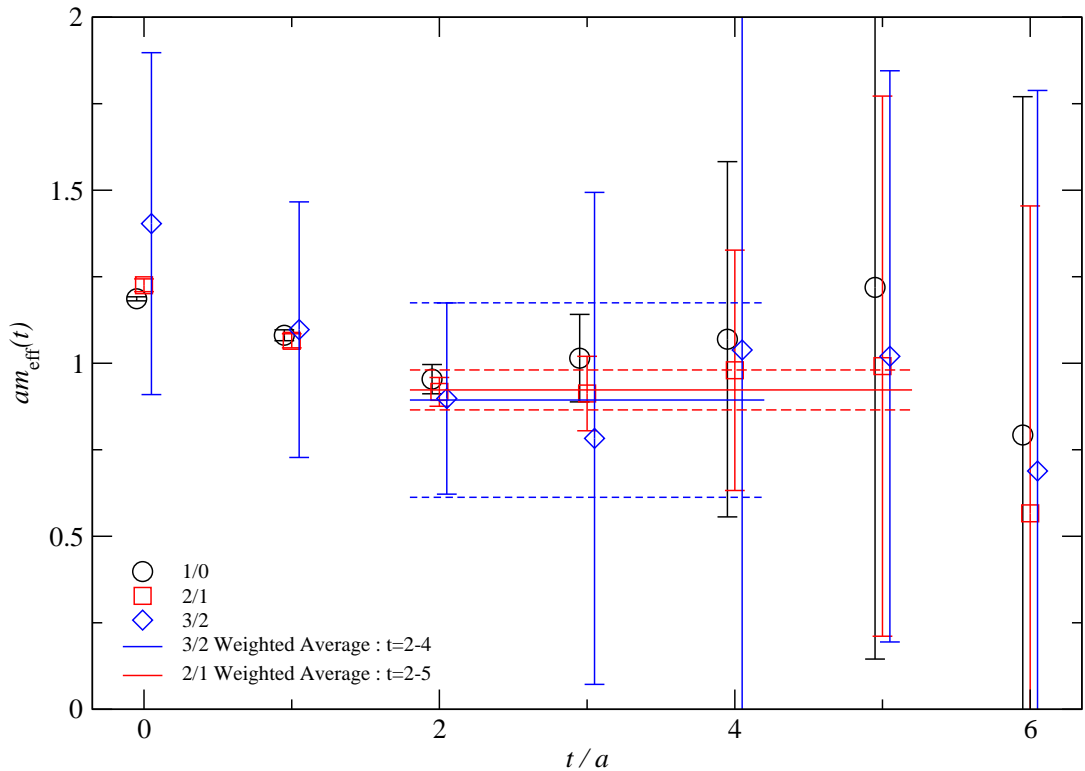}
\caption[Effective Mass (Coarse, Variational, Weighted
Average)]{Effective masses computed using the variational correlator 
for different choices of $t/t_0$.
Weighted averages are shown for the 
$2/1$ and $3/2$ projections (coarse ensemble, standard scalar 
glue operators).}
\label{fig:coarse_meff_var}
\end{figure}

Fig.~\ref{fig:coarse_meff_var} shows the variational effective masses,
as defined in section~\ref{sse:meas} (method 2b) 
deduced from a $3\times 3$ correlator matrix (blocking levels $0$ to $2$)

Similar effective mass estimates were made using momentum $1$
operators.  The latter require no vacuum subtraction.

\begin{figure}[ht]
\begin{center}
\includegraphics[width=0.9\textwidth,clip]
{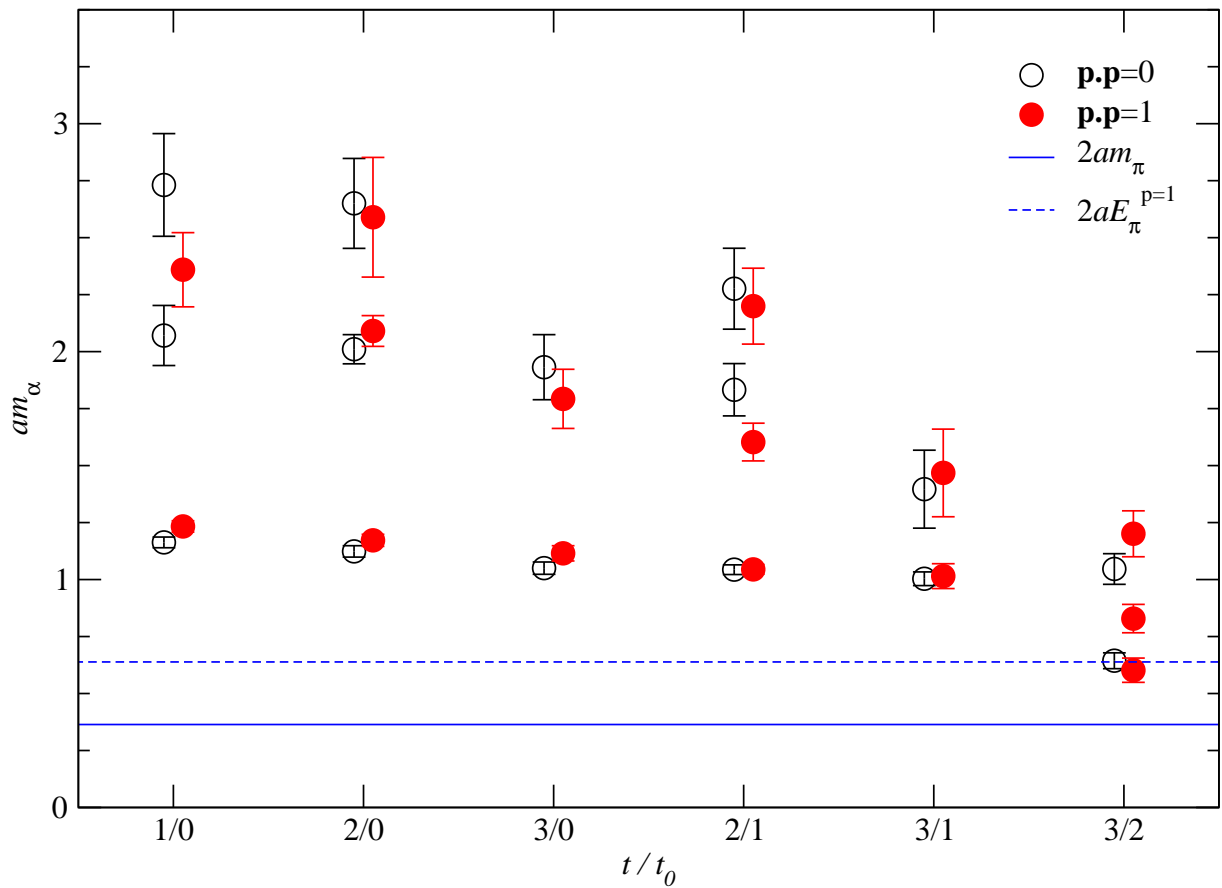}
\caption[Masses from the Variational Eigenvalues]
{The masses extracted from the variational
eigenvalues for different $t/t_0$ using 
blocking levels $0,2,3$ for momentum $0$ and $1,2,3$ for momentum $1$.
The energies corresponding to a $\pi\pi$ state with 
momentum $0$ and $1$ are drawn for comparison (coarse ensemble, 
standard scalar glue operators).}
\label{fig:coarse_vareig}
\end{center}
\end{figure}

In Fig.~\ref{fig:coarse_vareig} we show the lowest $3$ masses
(where obtainable) extracted from the variational matrix (method 2a).
The momentum $1$ masses have been obtained from energies using the
na\H{\i}ve lattice dispersion relation. 
There is a reasonable degree of consistency with respect to the 
choice of $t/t_0$
except at large $t$ where there is a suggestion that the
lowest state projected out corresponds to a $\pi\pi$ system. Given
that for these light quark masses, the two $\pi$ decay threshold is
open, this is not unexpected.

\begin{figure}[ht]
\begin{center}
\includegraphics[width=0.9\textwidth,clip]
{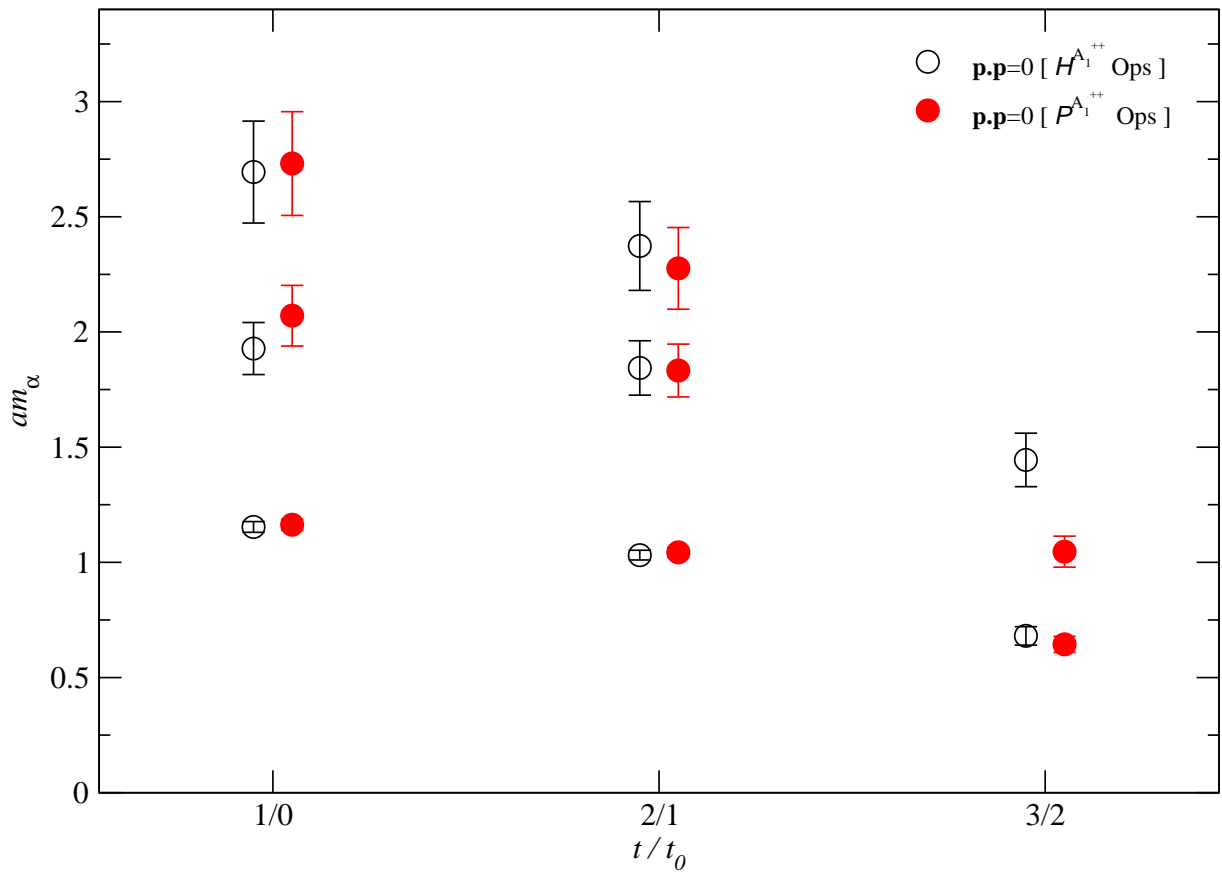}
\caption[Comparison of Masses obtained from the Variational
Eigenvalues with Different Ops (Coarse, Scalar)]{The masses extracted from 
  variational eigenvalues  for 
different $t/t_0$ (coarse ensemble). 
These were made using blocking levels
$0,1,2$ for standard plaquette (filled circles) and alternative hand 
(empty circles) operators.}
\label{fig:coarse_vareig_alt}
\end{center}
\end{figure}

Similar  studies were made using the alternative
hand operators defined in section~\ref{sse:meas}. 
The results were statistically consistent with those
obtained from the standard operators. Fig.~\ref{fig:coarse_vareig_alt}
shows a comparison of the lowest
three states using standard and alternative operators.

\begin{figure}[ht]
\begin{center}
\includegraphics[width=0.9\textwidth,clip]
{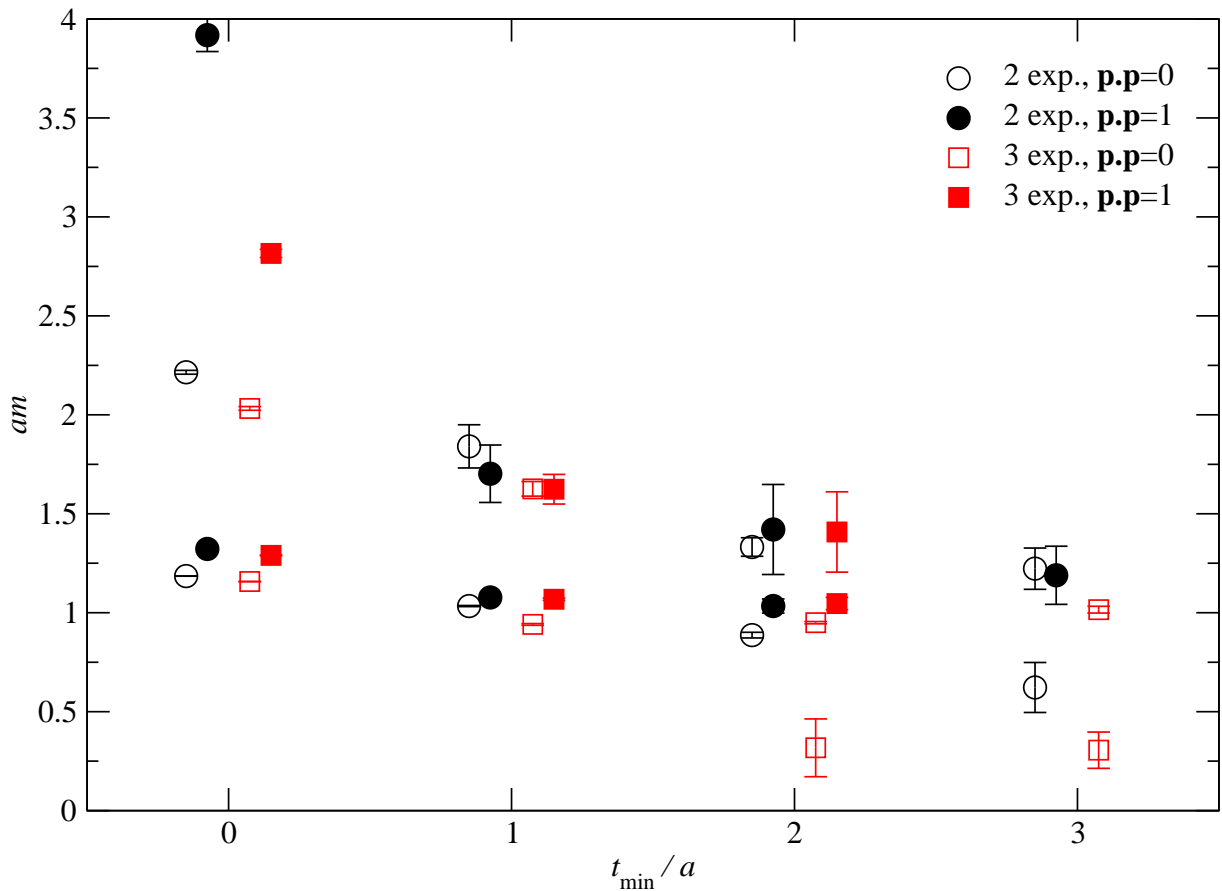}
\caption[Coarse Scalar Glueball Factorising Fits --- $4\times 4$
basis] {Results from factorising fits applied to the 
$4\times 4$ basis of standard scalar glueball operators 
with blocking levels $0$ to $3$.
Both two and three-exponential fits are 
presented for momentum $0$ and $1$
correlators, with $\tmax=6$ and $\tmin$ allowed to vary (coarse ensemble).}
\label{fig:coarse_facfit}\end{center}
\end{figure}

Finally in this subsection, we present results of the third method: multichannel
factorising fits. We have carried out $3\times 3$ and $4\times 4$ fits with $2$
and $3$ exponentials. We have studied both correlated and uncorrelated
fits. We also investigated stability with respect to the fitted $t$ range
$\tmin \leq t \leq \tmax $. 
For example, Fig.~\ref{fig:coarse_facfit} shows the variation with
$\tmin$ of the lowest two states resulting from two and three exponential fits to
a $4\times 4$ matrix of correlators (blocking levels $0-3$).
These fits are fully correlated and the corresponding fit details are
listed in Table~\ref{tab:coarse_facfit}.
\begin{table}[ht]
\centering
\begin{tabular}{|c|c|c|c|c|c|c|c|}
\hline
$N_\mathrm{exp}$ & $|p|$ & $\tmin$ & $\tmax$ & $am_0$ & $am_1$ &
$am_2$ & $\chisqpdof$ \\
\hline
 \multirow{3}{*}{2} & \multirow{3}{*}{0} & 1 & 6 & 1.0336(29) &
 1.84(11) & --- & $3.176$\\
 &  & 2 & 6 & 0.887(14) & 1.332(47) & --- & $0.635$\\
 &  & 3 & 6 & 0.62(13) & 1.22(10) & --- & $0.349$\\
\hline
 \multirow{3}{*}{2} & \multirow{3}{*}{1} & 1 & 6 & 1.0770(60) &
 1.70(14) & --- & 0.832 \\
 &  & 2 & 6 & 1.034(36) & 1.42(22) & --- & 0.301 \\
 &  & 3 & 6 & 1.19(15) & 3.30[$-$] & --- & 0.189 \\
\hline\hline
 \multirow{3}{*}{3} & \multirow{3}{*}{0} & 1 & 6 & 0.9409(43) &
 1.626(37) & 18.04[$-$] & 1.309 \\
 &  & 2 & 6 & 0.32(15) & 0.9495(57) & 1.77[$-$] & 0.607 \\
 &  & 3 & 6 & 0.306(92) & 1.016(17) & 7.79[$-$] & 0.327\\
\hline
  \multirow{2}{*}{3} & \multirow{2}{*}{1}& 1 & 6 & 1.0682(60) &
  1.623(75) & 21.26[$-$] & 0.329 \\
 & & 2 & 6 & 1.046(31) & 1.41(20) & 3.53[$-$] & 0.173 \\
\hline
\end{tabular}
\caption[Fit Results : Coarse Scalar Glueball ($4\times 4$)]
{Fitted mass parameters for two and three-exponential factorising 
fits to a $4\times 4$ matrix of correlators using blocking levels
$0-3$ (coarse ensemble).
Where errors are quoted as $[-]$ this indicates that the gradient 
in that direction of parameter space was undetermined.}
\label{tab:coarse_facfit}
\end{table}

\begin{figure}
\centering
\includegraphics[width=0.9\textwidth,clip]
{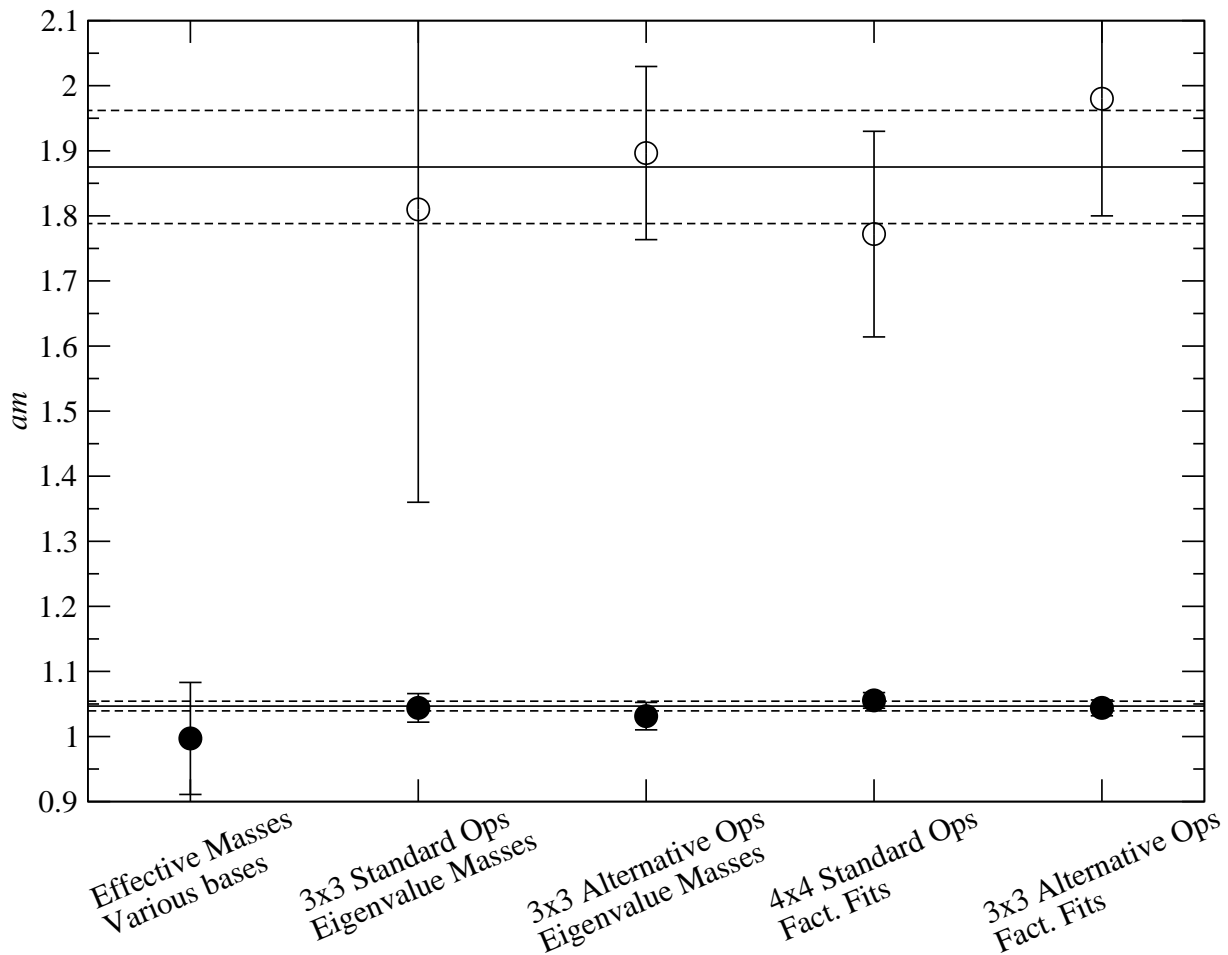}
\caption[Overall Mass Estimate for the Scalar Glueball (groundstate)
on the Coarse Lattices]{The overall average for the scalar glueball 
groundstate and first excited state masses on the coarse
ensemble, computed as described in the text.}
\label{fig:coarse_GSavg}
\end{figure}

The extent of consistency between the various methods can be judged
from Fig.~\ref{fig:coarse_GSavg}. 
The solid line, with dashed errors, represents a global average of
these determinations. This average was obtained using the methods
adopted by the Particle Data Group~\cite{Amsler:2008zzb} when
combining results from independent determinations. 
The errors used in the weighting are combined
systematic and statistical errors. Of course, the statistical errors
in our case are not independent. 
For each type of determination contributing to
Fig.~\ref{fig:coarse_GSavg}, represented by a central value and error
bar, we have used the same weighted averaging procedure.
The systematic errors took into account the variation due to fit range.
In fact, the determinations
contributing to the global average are a selected sub-group of all the
various estimates and fits described above~\cite{CMRT}. 
We have excluded fits
which had unacceptably high (or low) $\chisqpdof$ and/or which had very
large errors, and also those which were found to have high sensitivity
to fit ranges and/or parameter choices. Similar considerations were
applied to the determination of the first excited state in the 
scalar sector, also shown in Fig.~\ref{fig:coarse_GSavg}. 
We have only quoted fitted mass values for states
where at least one higher state has been included in the determination.
The global averages for the lowest two states in the scalar channel
are shown in Table~\ref{tab:ma_scalar}, along with the corresponding 
values for the fine lattices (described in the next subsection).
\begin{table}[!ht]
\begin{center}
\begin{tabular}{|l|c|c|c|}
\hline
ensemble & $a$ [fm] & $am$ & $am^*$ \\ 
\hline
coarse & 0.12250(24)  & $1.0468(75)$ & $1.875(87)$ \\
fine & 0.09230(19)  & $0.8332(59)$ & $1.368(17)$ \\
\hline
\end{tabular}
\caption{Global average values for scalar masses on coarse and 
fine ensembles}
\label{tab:ma_scalar}
\end{center}
\end{table}
We will return later to the physical interpretation of these results.

\subsection{Results for the scalar glueball -- fine ensemble}\label{sse:fine}
\begin{figure}[ht]
\begin{center}
\includegraphics[width=0.8\textwidth,clip]
{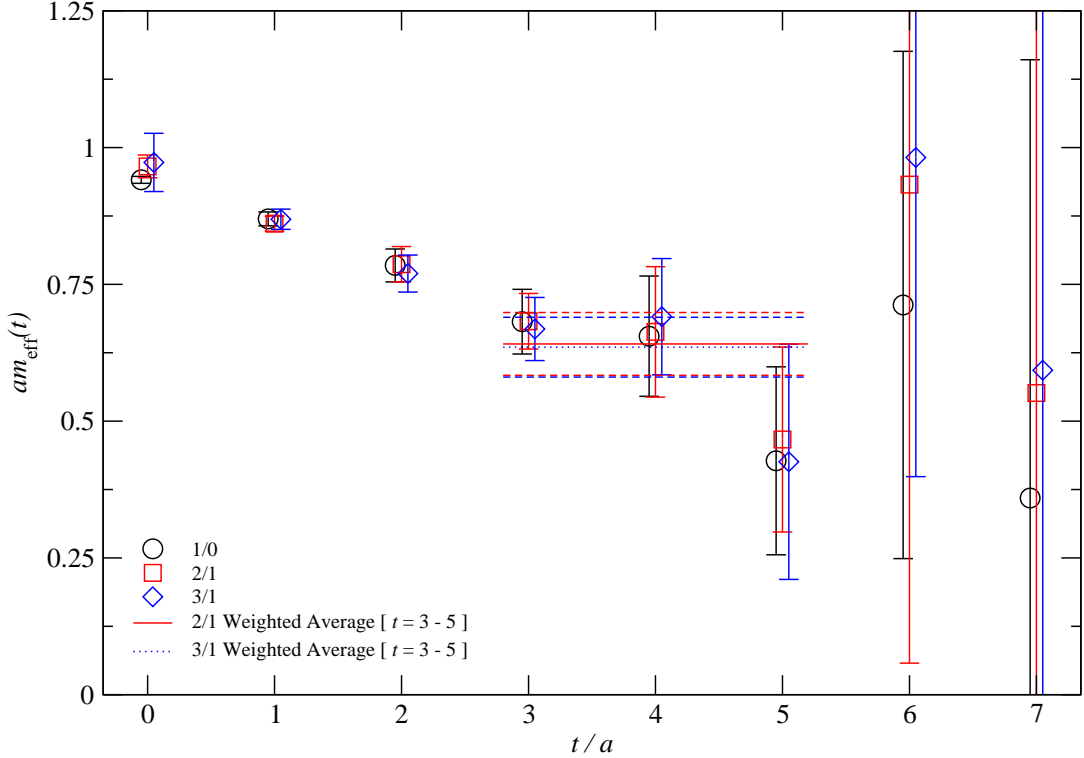}
\caption[Effective Mass: Fine, Variational]
{Variational effective masses for the scalar glueball 
computed with different choices of
$t/t_0$, projecting from a $4\times 4$ matrix (blocking levels $0,1,2,3$)
measured on the fine lattices.
Weighted averages are shown for the $2/1$ and $3/1$ projections.}
\label{fig:fine_meff_var}
\end{center}
\end{figure}

\begin{figure}[ht]
\begin{center}
\includegraphics[width=0.9\textwidth,clip]
{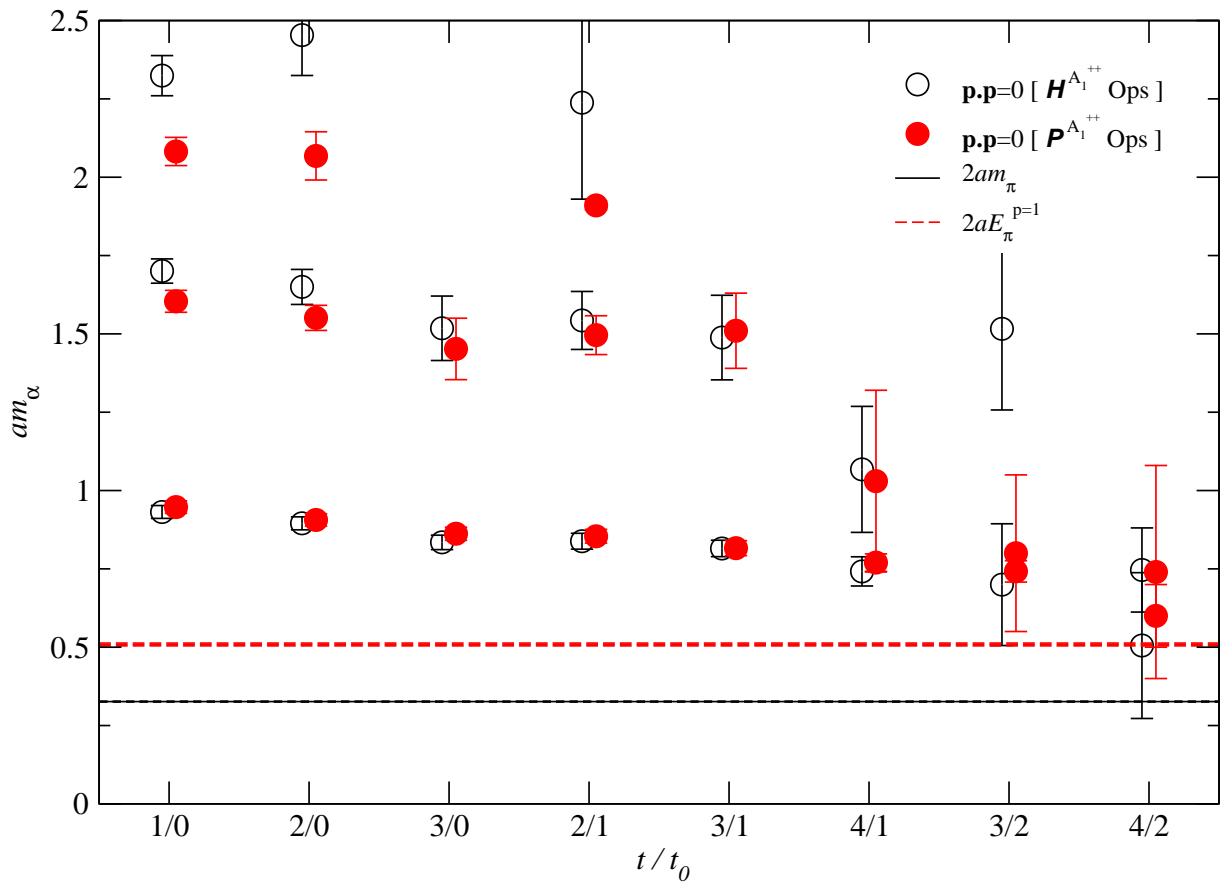}
\caption[Masses from the Variational Eigenvalues ]
{Scalar glueball masses extracted, for the fine ensemble, 
from the variational eigenvalues using
blocking levels $0,1,2$ for different $t/t_0$. 
Results based on the standard and alternative (hand) operators are compared.
The energies corresponding to
a $\pi\pi$ state with relative momenta $|p|=0$ and $|p|=1$ 
are drawn for comparison (fine ensemble).}
\label{fig:fine_vareig}
\end{center}
\end{figure}

Fig.~\ref{fig:fine_meff_var} (corresponding to Fig.~\ref{fig:coarse_meff_var} 
for the coarse ensemble) shows 
variational effective scalar masses and weighted averages on the fine
ensemble.
In Fig.~\ref{fig:fine_vareig}, we show the lowest 3 masses extracted from the
variational matrix as described in method 2a. 
This is to be compared
with Fig.~\ref{fig:coarse_vareig} for the coarse ensemble, but in this
case we take the opportunity to display results for both standard
and alternative (hand) operators. 
Both operator bases gave consistent results.

\begin{figure}[ht]
\begin{center}
\includegraphics[width=0.9\textwidth,clip]{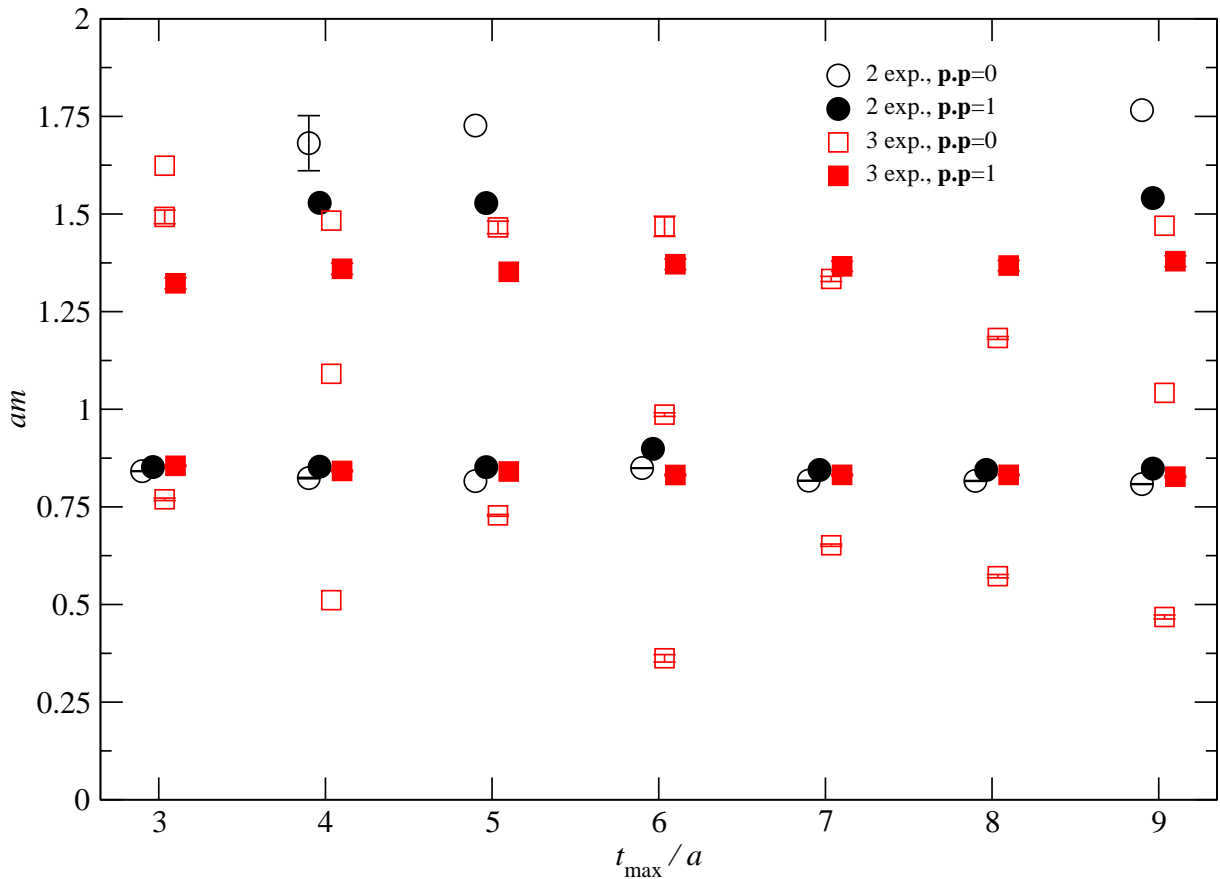}
\caption[Fine Scalar Glueball Factorising Fits]
{Results from factorising fits applied to the $4\times 4$ basis 
of standard scalar glueball operators (blocking levels $0,1,2,3$)
measured on the fine lattices. Both two and three-exponential fits 
are presented for momentum $0,1$ correlators,
with $\tmin=1$ and $\tmax$ allowed to vary.}
\label{fig:fine_facfit}
\end{center}
\end{figure}

As examples of factorising fits on the fine ensemble, 
we present the $\tmax$ dependence at fixed $\tmin$ in 
Fig.~\ref{fig:fine_facfit}. These are $4\times 4$ fits using two and
three exponentials. 
We see excellent consistency for the groundstates between 
the $M=2$ (both $|p|=0$ and 1) and the $M=3$,
$|p|=1$ fits. 
The corresponding fit quality can be judged from
Tables~\ref{tab:fine_facfit2} and~\ref{tab:fine_facfit3}.
\begin{table}[ht]
\centering
\begin{tabular}{|c|c|c|c|c|c|c|c|}
\hline
$N_\mathrm{exp}$ & $|p|$ & $\tmin$ & $\tmax$ & $am_0$ & $am_1$ &
$am_2$ & $\chisqpdof$ \\
\hline
\multirow{7}{*}{2} & \multirow{7}{*}{0} & 1 & 3 & 0.84139(97) &
19.84[$-$] & --- & 7.233 \\
 &  & 1 & 4 & 0.8237(17) & 1.681(71) & --- & 5.098 \\
 &  & 1 & 5 & 0.82[$-$] & 1.73[$-$] & --- & 5.255 \\
 &  & 1 & 6 & 0.84937(76) & 692[$-$] & --- & 31.42 \\
 &  & 1 & 7 & 0.81705(91) & 19.16[$-$] & --- & 6.747 \\
 &  & 1 & 8 & 0.81657(92) & 19.06[$-$] & --- & 6.515 \\
 &  & 1 & 9 & 0.80834(81) & 1.77[$-$] & --- & 5.182 \\
\hline
\multirow{7}{*}{2} & \multirow{7}{*}{1} & 1 & 3 & 0.8521(15) &
19.56[$-$] & --- & 8.185 \\
 &  & 1 & 4 & 0.8528(13) & 1.53[$-$] & --- & 3.604 \\
 &  & 1 & 5 & 0.8517(12) & 1.53[$-$] & --- & 2.885 \\
 &  & 1 & 6 & 0.8985(14) & 692[$-$] & --- & 37.12 \\
 &  & 1 & 7 & 0.8446(13) & 19.00[$-$] & --- & 3.838 \\
 &  & 1 & 8 & 0.8446(13) & 18.93[$-$] & --- & 3.368 \\
 &  & 1 & 9 & 0.8480(11) & 1.54[$-$] & --- & 1.933 \\
\hline
\end{tabular}
\caption[Fit Results : Fine Scalar Glueball ]{Fitted mass 
parameters for two-exponential fully correlated
factorising fits to a $4\times 4$ matrix of correlators 
using a blocking levels $0,1,2,3$
 with
$|p|=0$ and $1$
for varying $\tmax$ with $\tmin$ fixed at $t=1$
(fine ensemble). Where
errors are quoted as $[-]$ this indicates that the 
gradient in that direction of parameter 
space could not be determined.}
\label{tab:fine_facfit2}
\end{table}
\begin{table}[ht]
\centering
\begin{tabular}{|c|c|c|c|c|c|c|c|}
\hline
$N_\mathrm{exp}$ & $|p|$ & $\tmin$ & $\tmax$ & $am_0$ & $am_1$ &
$am_2$ & $\chisqpdof$ \\
\hline
\multirow{7}{*}{3} & \multirow{7}{*}{0} & 1 & 3 & 0.7695(31) &
1.493(17) & 1.62[$-$] & 1.549 \\
 &  & 1 & 4 & 0.51[$-$] & 1.09[$-$] & 1.48[$-$] & 1.450 \\
 &  & 1 & 5 & 0.7284(22) & 1.466(16) & 20.29[$-$] & 1.987 \\
 &  & 1 & 6 & 0.3622(96) & 0.9863(45) & 1.468(26) & 1.126 \\
 &  & 1 & 7 & 0.6518(30) & 1.3337(66) & 16.65[$-$] & 2.495 \\
 &  & 1 & 8 & 0.5726(43) & 1.1827(37) & 19.78[$-$] & 2.648 \\
 &  & 1 & 9 & 0.4679(51) & 1.04[$-$] & 1.47[$-$] & 1.866 \\
\hline
\multirow{7}{*}{3} & \multirow{7}{*}{1} & 1 & 3 & 0.8551(21) &
1.322(14) & 2.67(12) & 0.144 \\
 &  & 1 & 4 & 0.8420(21) & 1.360(14) & 2.73(20) & 0.431 \\
 &  & 1 & 5 & 0.84[$-$] & 1.352(13) & 22.23[$-$] & 0.378 \\
 &  & 1 & 6 & 0.8317(22) & 1.371(14) & 2.57[$-$] & 0.542 \\
 &  & 1 & 7 & 0.8319(20) & 1.366(13) & 18.17[$-$] & 0.642 \\
 &  & 1 & 8 & 0.8319(20) & 1.368(13) & 21.12[$-$] & 0.585 \\
 &  & 1 & 9 & 0.8274(23) & 1.379(14) & 2.40[$-$] & 0.540 \\
\hline
\end{tabular}
\caption[Fit Results : Fine Scalar Glueball ]{Fitted mass 
parameters for three-exponential fully correlated
factorising fits to a $4\times 4$ matrix of correlators 
using a blocking levels $0,1,2,3$ 
 with
$|p|=0$ and $1$
for varying $\tmax$ with $\tmin$ fixed at $t=1$ (fine ensemble).}
\label{tab:fine_facfit3}
\end{table}

\begin{figure}[ht]
\begin{center}
\includegraphics[width=0.9\textwidth,clip]{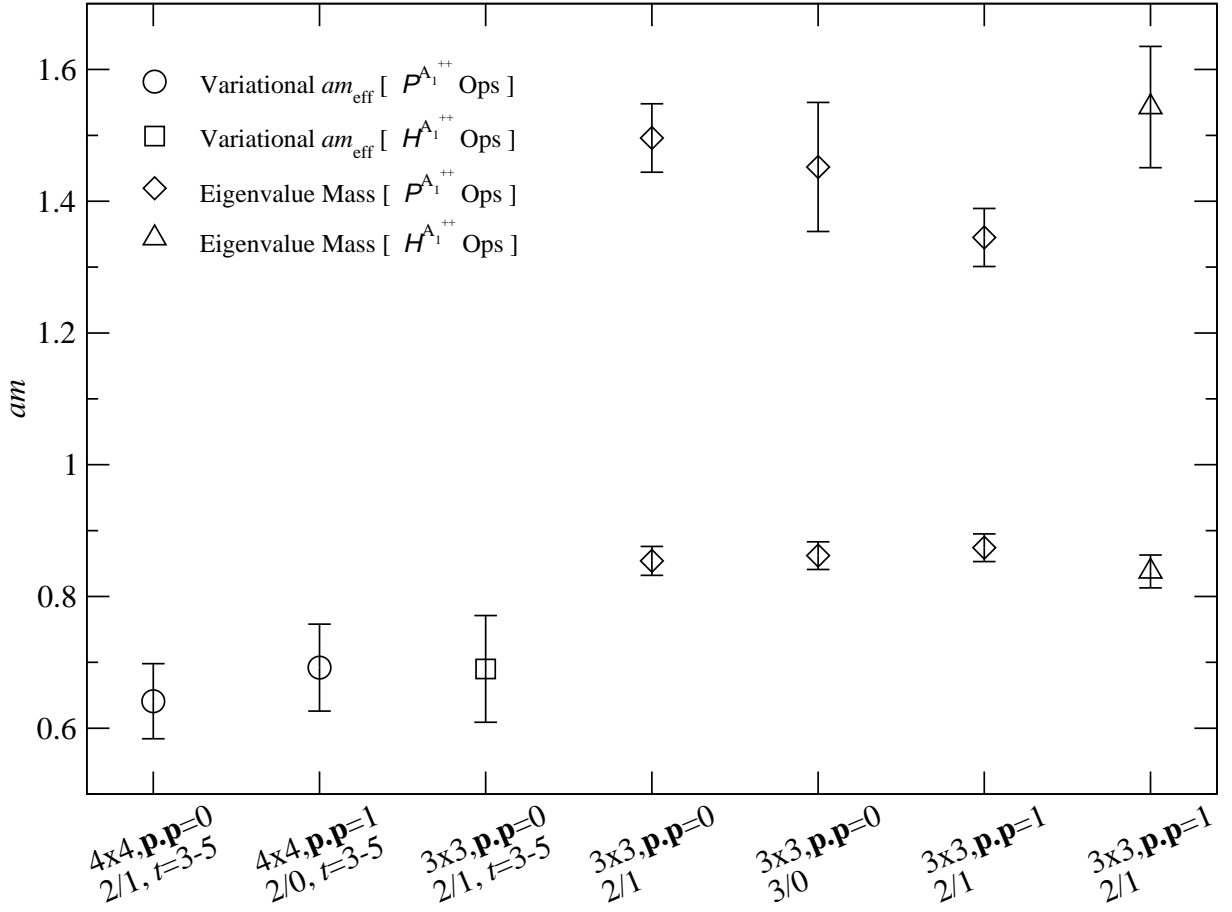}
\caption[Comparison of Masses obtained for the Scalar Glueball on the
Fine Lattices using different methods]
{Comparison of mass estimates obtained for the scalar glueball using
  different methods (variational effective masses and variational eigenvalues) for
  the fine ensemble. 
The points are described briefly on the axis and in the text.}
\label{fig:mass_compare_scalar_fine_meff}
\end{center}
\end{figure}

As with the coarse ensemble, we have selected the most reliable
determinations for each of the described methods and assigned
systematic errors to cover sensitivity to parameter and fit range.
As an example, Fig.~\ref{fig:mass_compare_scalar_fine_meff}
shows sample variational results (effective mass and eigenvalue
methods) for momentum zero and one. The relevant time slices $t/t_0$
are shown on the horizontal axis along with the $t$ range used for
averaging the effective mass.

\begin{figure}[ht]
\begin{center}
\includegraphics[width=0.9\textwidth,clip]{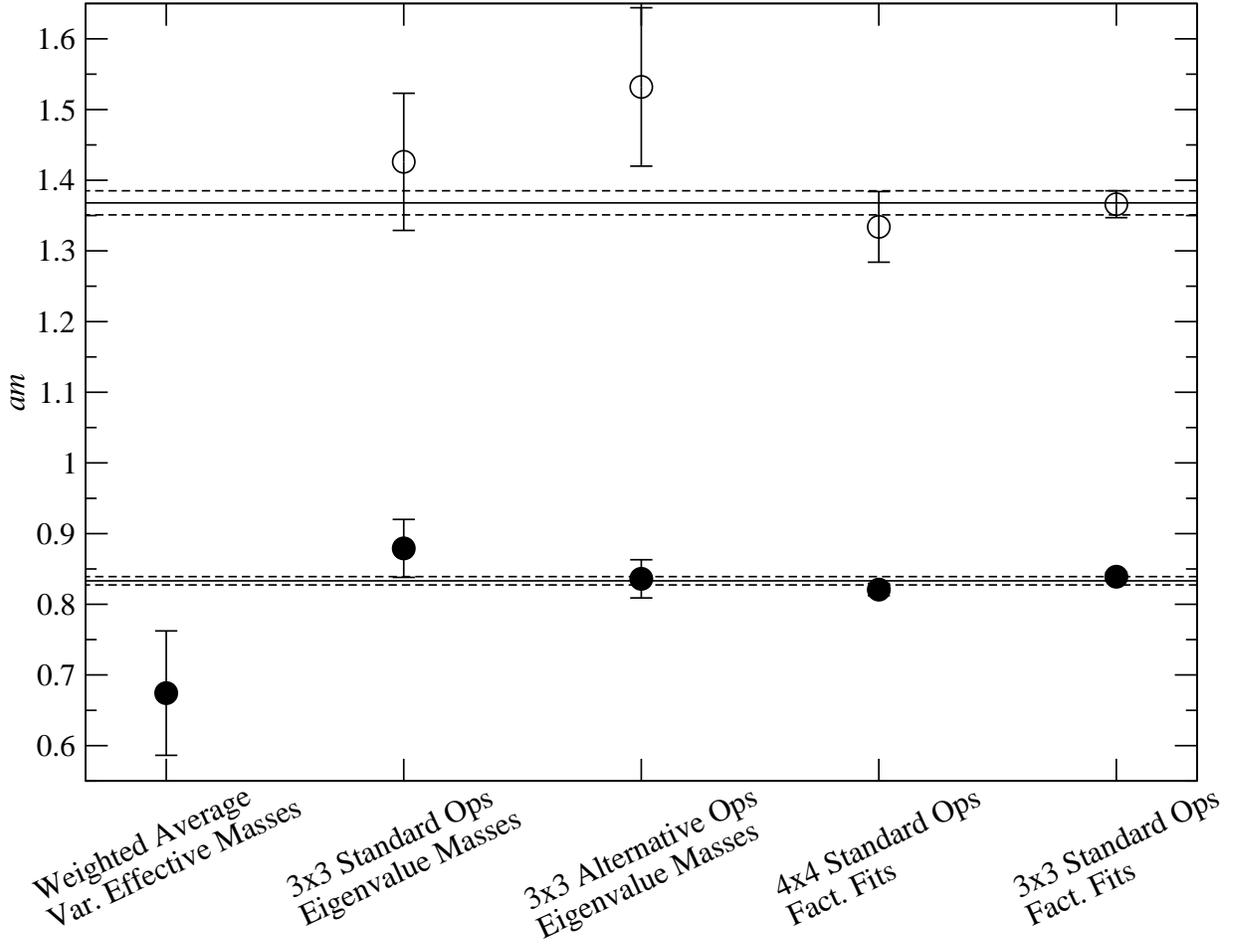}
\caption[Overall Mass Estimate for the Scalar Glueball (groundstate)
on the Fine \
Lattices]{The overall average for the scalar glueball groundstate mass
  on the fine ensemble, evaluated as described in the text.}
\label{fig:fine_GSavg}
\end{center}
\end{figure}

The final values selected for all 
methods are shown in Fig.~\ref{fig:fine_GSavg} along with the
resulting global average using the same weighting procedures
described earlier in section~\ref{sse:coarse}.

\subsection{Scalar decay and mixing}\label{sse:decay}
Having observed signals in our fits which appear close to the $2\pi$
threshold, we have attempted to check the strength of mixing
between the glueball and $\pi\pi$ operators. 
Early attempts to study the mixing directly have been 
made \cite{Sexton:1995kd} but were performed at pion
masses such that $2m_\pi$ was just below $m_G$. 
One might na\H{\i}vely expect that glueball decay 
would be flavour blind, decaying into $\pi\pi$,~$K\bar{K}$ and
$\eta\eta$ final states with equal rates. 
Of course, since we are now observing the state (or states) 
to which the gluonic operators couple most strongly, then mixing
effects are likely to violate flavour blindness.
If the `glueball' mixes strongly with a mainly $\uubar+\ddbar$ 
state then the $\pi\pi$ decay channel is preferred
over the OZI suppressed $K\bar{K}$ channel, and vice versa if the coupling is
strongest to an $\ssbar$ state.
There are chiral suppression arguments~\cite{Chanowitz:2005du}
that suggest that KKbar could dominate the two meson decay
channel in a glueball state in this mass range.
 
\begin{figure}[!htbp]
\begin{center}
\includegraphics[width=0.9\textwidth,clip]{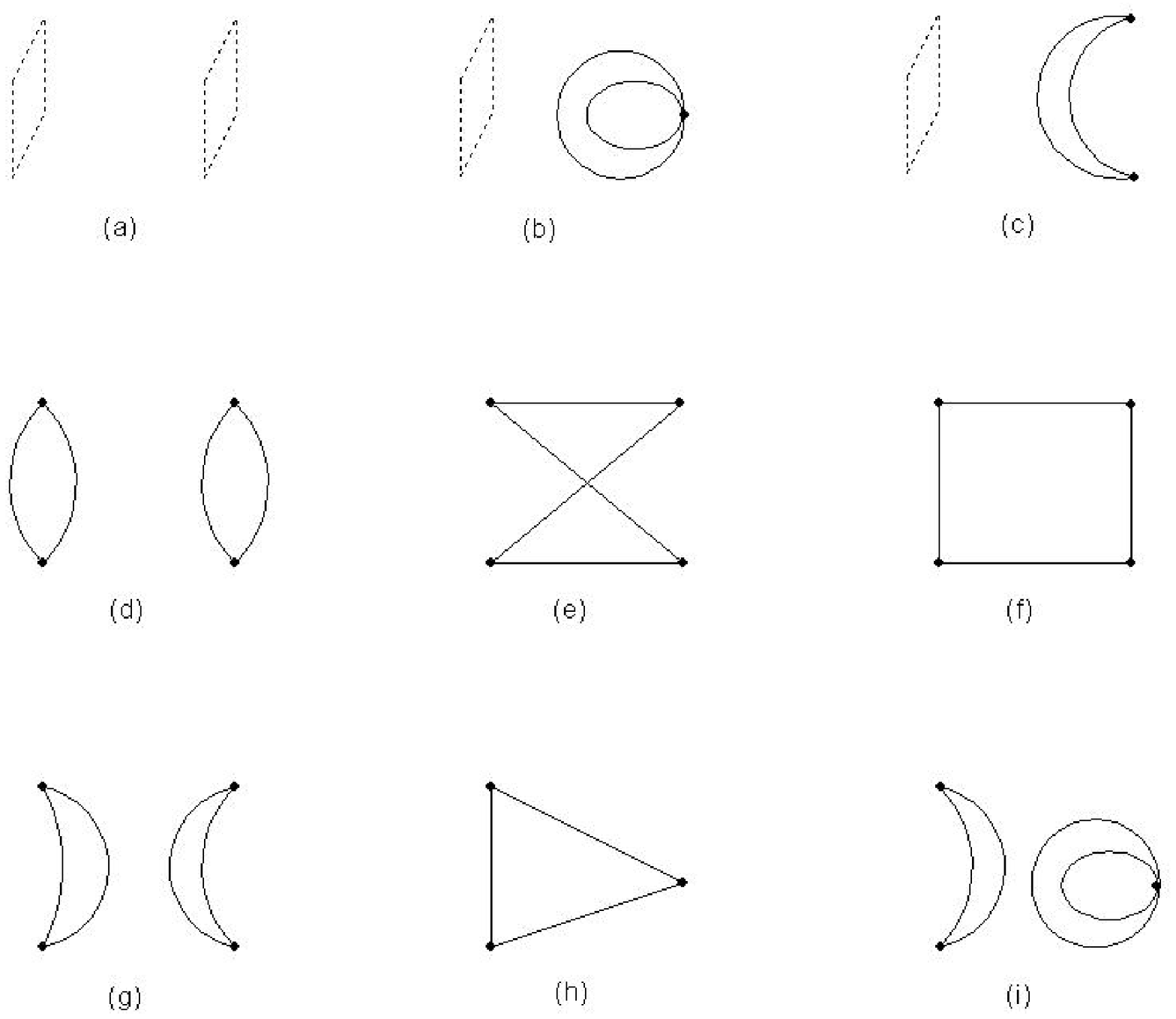}
\caption[Schematic of possible mixing diagrams]
{Schematic mixing diagrams for the full glue, meson, and two meson mixing
 problem in the $I=0$ scalar channel. The dashed rectangles represent
 the glue operators $\Pop^{\App}$ and the solid lines
 represent quark propagators. Euclidean time runs horizontally.}
\label{fig:mixing}
\end{center}
\end{figure}

One can in principle study decay matrix elements directly on the
lattice following the procedure of 
Lellouch and L\"uscher \cite{Lellouch:2000pv}.
However this requires very accurate determinations 
on different lattice sizes and is a difficult 
technique to apply even for states 
which are much less subject to noise 
than those in the glueball sector.
Initial attempts to study resonant states with open decay channels
have been made - for example the PACS-CS 
collaboration~\cite{Aoki:2007rd}
has studied the isovector $P$-wave phase shift in $\pi\pi$
scattering in order to estimate the width of the $\rho$
meson. Additional complications and extra diagrams introduced in the
case of staggered fermions have been detailed for example 
in~\cite{Sharpe:1992pp}
where $I=2$ phase shifts were studied.
In the isoscalar channel there are of course the further complications
arising from disconnected contributions. The possible types of
correlators required for a mixing study of glueball, $f_0$ and
$\pi\pi$ are indicated in Fig.~\ref{fig:mixing}.
Unquenched studies of mixing in the isoscalar $0^{++}$ system
were carried out by Hart et al.~\cite{Hart:2006ps}
using two flavours of Wilson fermions. 
The ratio of diagram (i) to to diagram (h) (see Fig.~\ref{fig:mixing})
was estimated using all to all techniques with 100
configurations. Unfortunatly the errors were too large to provide
useful information within the statistics of that study.

It has become clear that techniques for studying
lattice decays will best be developed first in 
simpler and less noisy systems than isoscalar channels
such as the present one.
  
So, for a preliminary study with improved
staggered fermions, we 
choose instead to follow a simplified version of the 
method of~\cite{McNeile:2000xx} and form the mixing matrix
\begin{equation}
\label{eq:gb_pipi_mix}
M_{ab}(t)=\left(\begin{array}{cc}
M_{GG}(t)&M_{\pi G}(t)\\
M_{G\pi}(t)&M_{\pi\pi}(t)\end{array}\right)=
\left(\begin{array}{cc}
\langle \Pop^{\App}_i(0)\Pop^{\App}_i(t)\rangle & \langle
C_\pi(0)\Pop^{\App}_i(t)\rangle \\
\langle \Pop^{\App}_i(t)C_\pi(0)\rangle & \langle
C_\pi(t)C_\pi(t)\rangle\end{array}
\right)
\end{equation}
where $\Pop^{\App}_i(0)$ and $\Pop^{\App}_i(t)$ are the $\modp=0$
standard scalar glueball operators at timeslices $0$ and $t$
respectively, and $C_\pi(0)$ and $C_\pi(t)$ are the single pion correlators on
timeslice $0$ and  $t$ respectively. 
The $C_\pi(0)$ correlator allows us to study a $\pi\pi$ state 
localised in time and, because of the way we have computed the connected
correlator, we are restricted to using $C_\pi(0)$ 
which reduces our statistics for the $2\pi$  operators by a factor 
of $n_t=64$. 

We form the ratio \cite{McNeile:2000xx}
\begin{equation}
\label{eq:gb_pipi_mix_rat}
x_{G\pi}(t)=\frac{M_{G\pi}(t)}{\sqrt{M_{GG}(t)M_{\pi\pi}(t)}}
\end{equation}
where the $M_{ab}(t)$ are the elements of (\ref{eq:gb_pipi_mix})
on timeslice $t$. 
This gives a measure of the off-diagonal mixing matrix
elements, normalised by the diagonal entries.
We remind the reader that this captures only some of the possible
contractions contributing in a full treatment of the mixing as
indicated in Fig.~\ref{fig:mixing}. In this study, the factorising
fits involved diagrams of type (a) only and the estimated
strength of mixing to
$\pi\pi$ involved an approximation to (c).

\begin{figure}[ht]
\centering
\subfigure[Coarse: LL Pion]{\label{fig:coarse_gbpi_ratio_LL_3}
 \includegraphics[width=0.45\textwidth,clip]{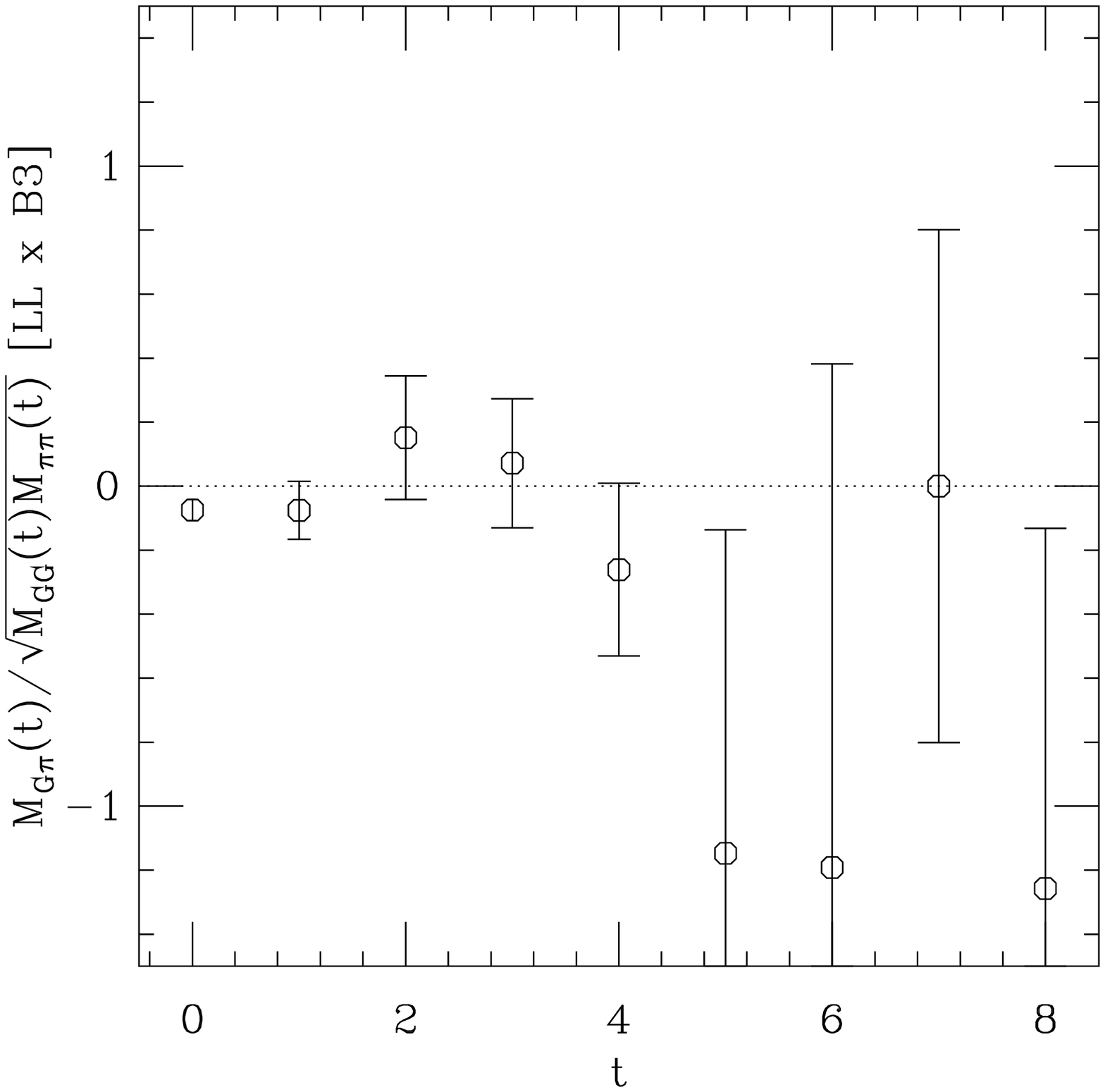}}
\subfigure[Coarse: FF Pion]{\label{fig:coarse_gbpi_ratio_FF_3}
 \includegraphics[width=0.45\textwidth,clip]{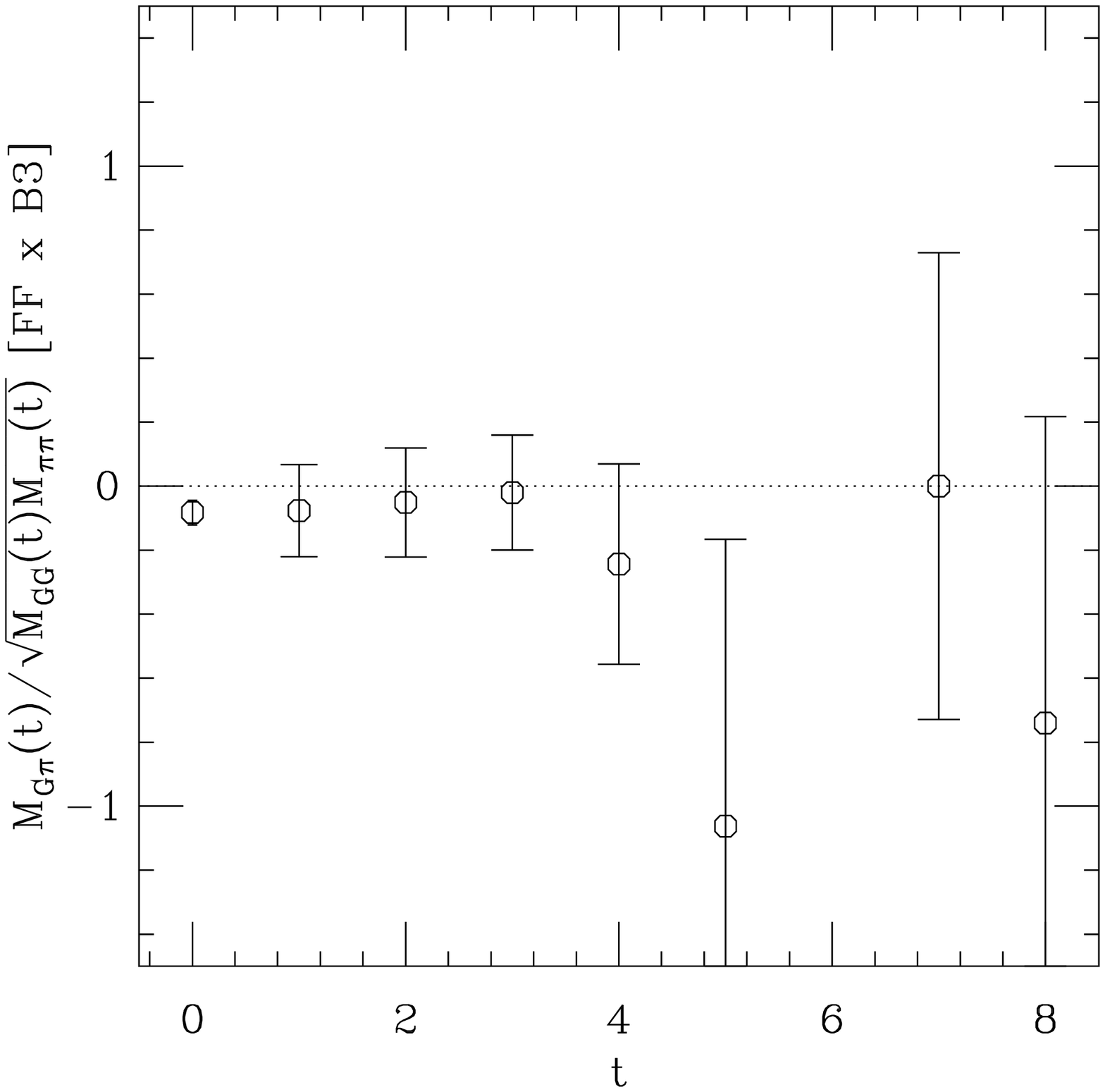}}
\subfigure[Fine: LL Pion]{\label{fig:fine_gbpi_ratio_LL_3}
 \includegraphics[width=0.45\textwidth,clip]{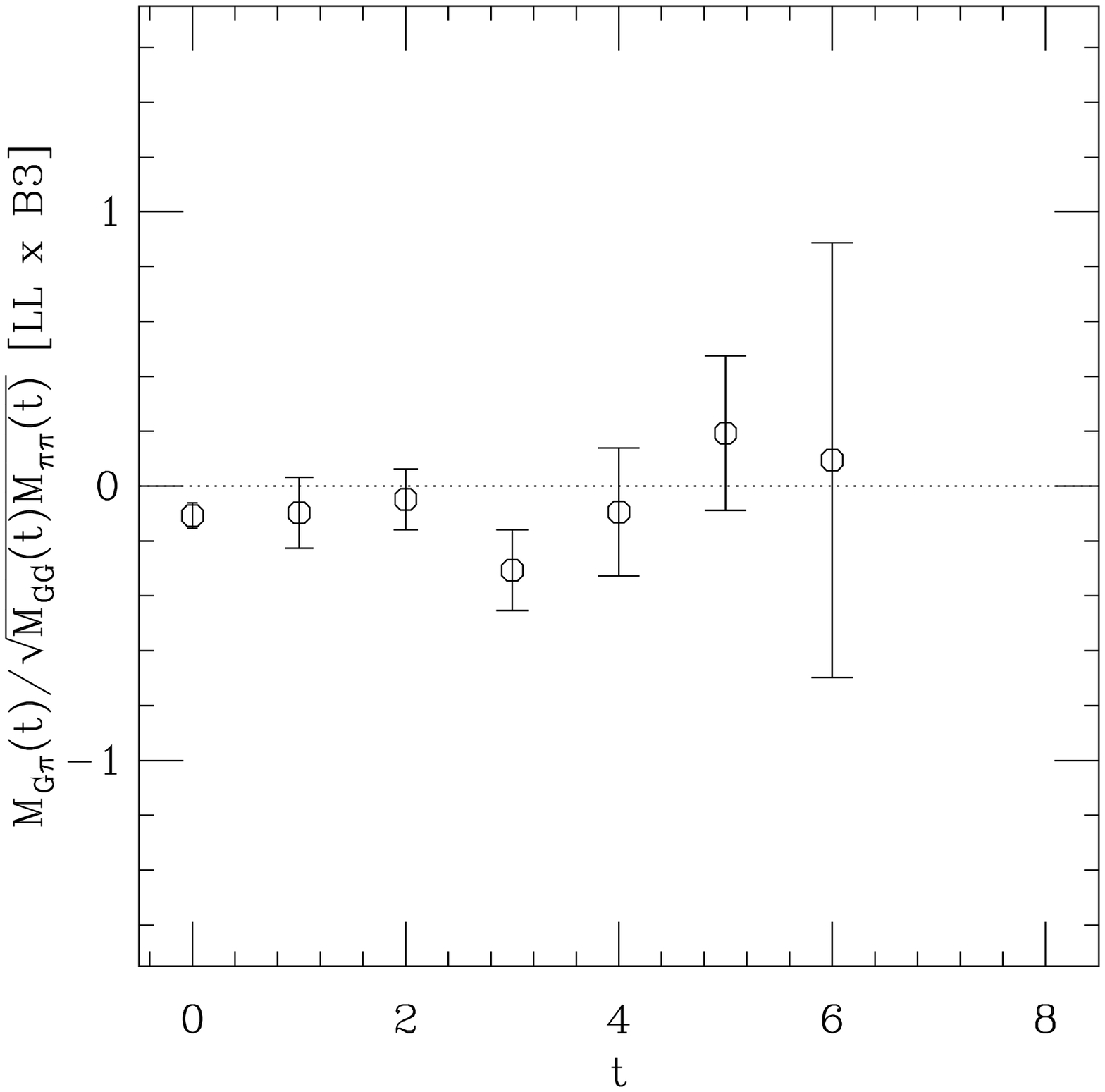}}
\subfigure[Fine: FF Pion]{\label{fig:fine_gbpi_ratio_FF_3}
 \includegraphics[width=0.45\textwidth,clip]{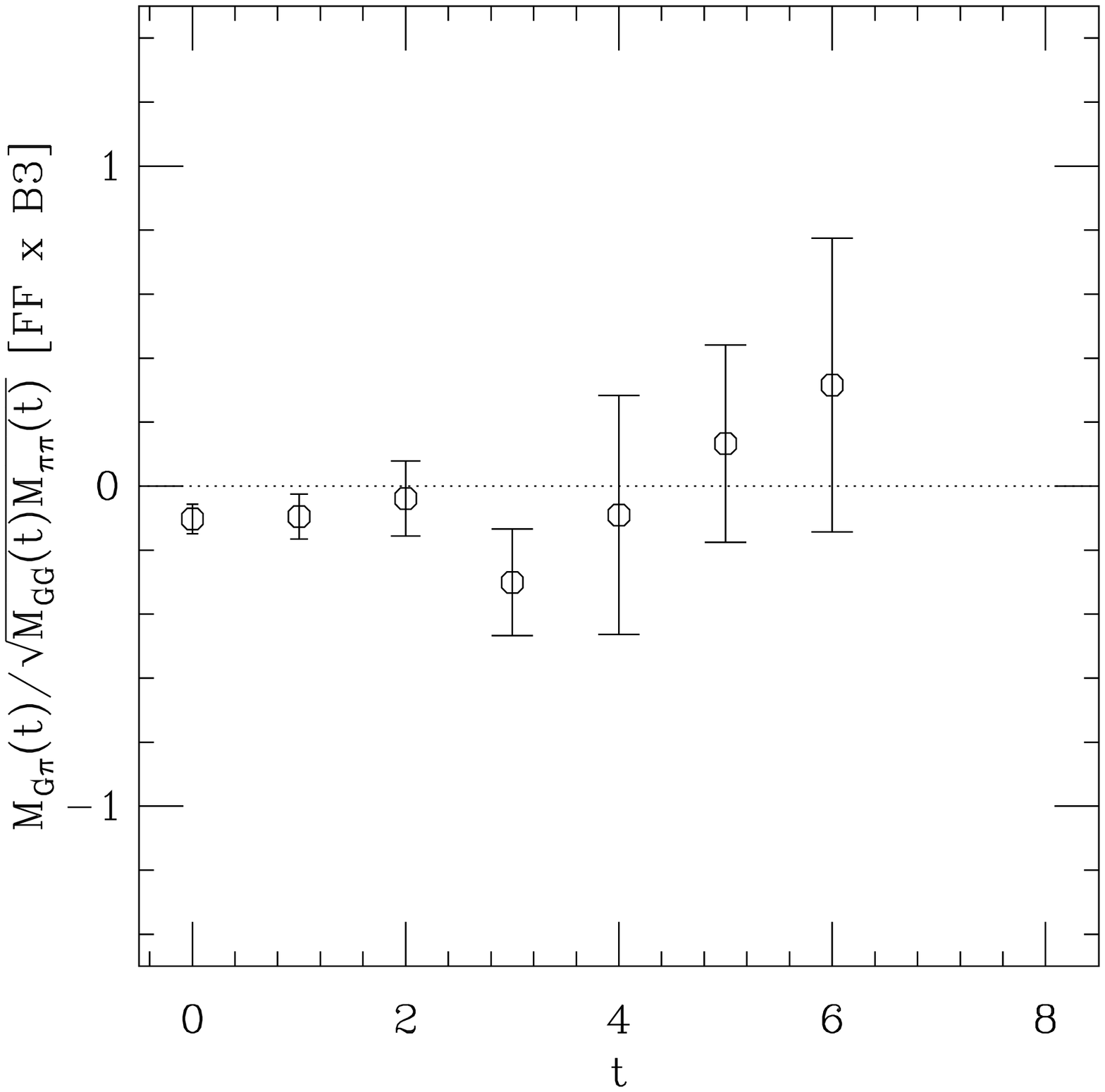}}
\caption[Glueball-$\pi\pi$ mixing for the Coarse and Fine lattices
with the $LL$ and $FF$ Pion Correlators]
{The measure of glueball-$\pi\pi$ mixing
  defined in (\ref{eq:gb_pipi_mix_rat}) for, 
clockwise from top-left: the local-local (LL) pion correlators 
with the three-times Teper blocked glueball operators on the
coarse ensemble; the fuzzed-fuzzed (FF) pion correlators with the three-times
blocked glueball operators on the coarse ensemble; the fuzzed-fuzzed (FF)
 pion correlators with the three-times 
blocked glueball operators on the fine ensemble; and 
the local-local (LL) pion correlators with the three-times 
blocked glueball operators on the fine ensemble.}
\label{fig:gbpi_ratios}
\end{figure}

We have computed these ratios for the coarse and fine lattices, using
the glueball operator with three levels of 
Teper blocking (\ie $\Pop_3^{\App}$) which we found 
shows similar behaviour to the lower blocking levels but with
less noise.
We used the pion correlators with both local source and local sink,
and fuzzed source and fuzzed sink. 
Our results are presented in Fig.~\ref{fig:gbpi_ratios}.

For the coarse ratios we see that at small $t$ they appear consistent
with zero, turning negative for $t\sim 5$, 
although with large error. 
We note that the fine mixing ratio shows a similar downturn
for $t\sim\ 3-4$. 
Whilst these ratios give only a guide to the mixing of the
glueball operators and a $\pi\pi$ state, the small size 
of $x_{G\pi}$ at low $t$
indicates that, provided we choose $\tmin$ small enough, 
we should obtain a good overlap with the \lq stable\rq{} glueball 
and conversely, that by choosing $\tmin$ large, we may
obtain a significant overlap with the $\pi\pi$ state. 
This is consistent with the earlier observations 
of the effective masses
and factorising fits at small and large $t$.

\subsection{Comparison with other scalar determinations}
\label{sse:scalar_comparisons}
This study of scalar states in the glueball sector is 
characterised by relatively high statistics in comparison with
previous studies using dynamical fermions. 
It also uses a vacuum with $2+1$ flavours. 
However, it is limited in that it makes
use of only two lattice spacings (roughly $0.12$ and $0.09$ fm) and
a single light quark mass for each ensemble. The pion mass in lattice
units is $0.1740(6)$ (coarse ensemble) and $0.1672(14)$ (fine).
These correspond to around $280$ MeV and $360$ MeV respectively so the
quark masses are not particularly light. The strange quark mass is
close to the physical one.
Any attempt to extract a continuum limit  
(without a chiral limit of course) needs to be treated with due
caution. In Table~\ref{tab:gbmasses_in_r0} we summarise the scalar sector
results in lattice units and in dimensionless form using the
Sommer $r_0$ parameter from Table~\ref{tab:ensparams}.
\begin{table}[ht]
\centering
\begin{tabular}{|r|c|c|}
\hline
Result & $am(\App)$ & $r_{0}m(\App)$\\
\hline
Coarse -- Ground & 1.0468(75) & 3.991(36)\\
Fine -- Ground & 0.8332(59) & 4.215(38) \\
Coarse -- Excited & 1.875(87) & 7.15(35) \\
Fine -- Excited & 1.368(17) & 6.92(10) \\
\hline
\end{tabular}
\caption[Scalar Glueball Masses in Units of $r_0$]
{Scalar glueball masses (ground and first-excited states) 
from the coarse and fine lattices converted
 into units of the Sommer parameter $r_0$.}
\label{tab:gbmasses_in_r0}
\end{table}

\begin{figure}[ht]
\begin{center}
\includegraphics[width=0.9\textwidth,clip]{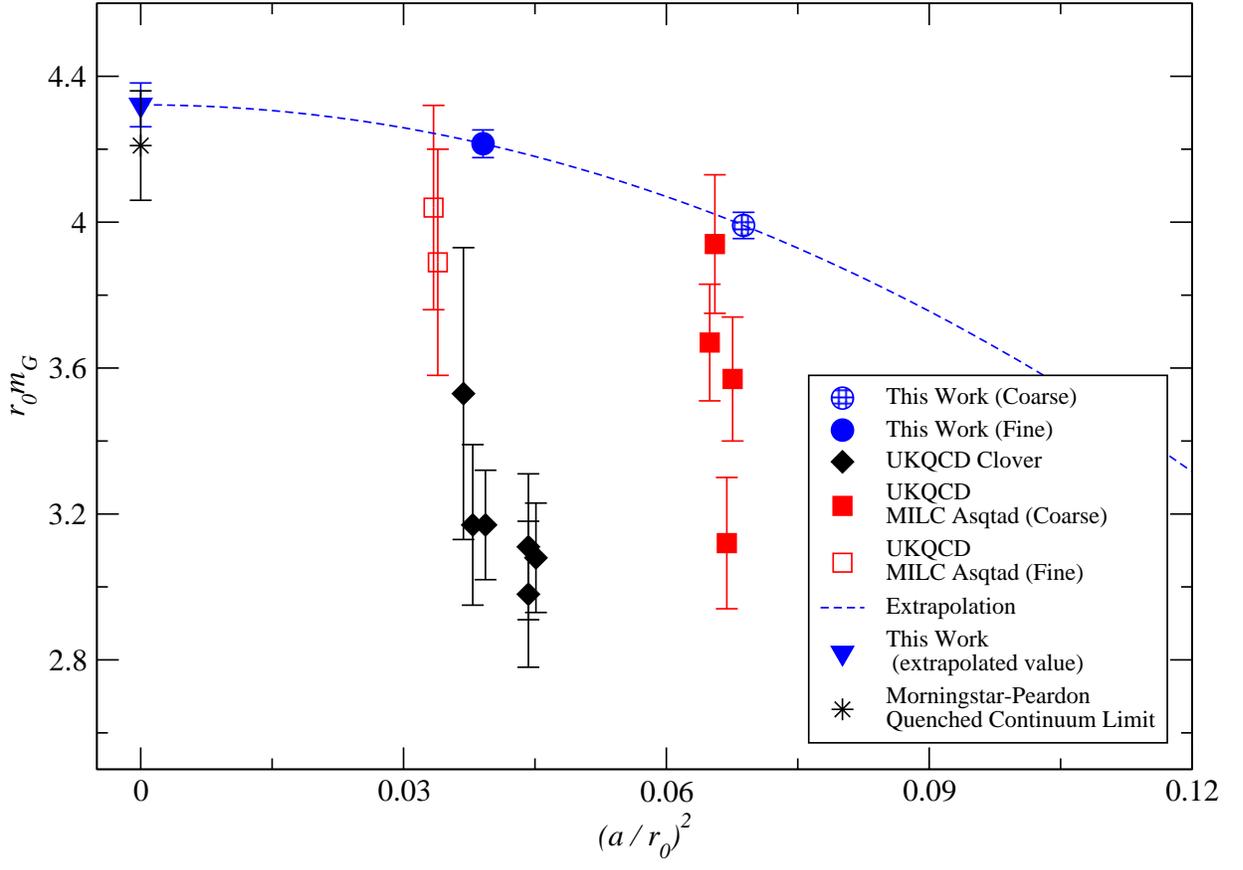}
\caption[Comparison of Scalar Glueball Masses with Previous
Determinations]
{Our measurements of the scalar glueball mass shown with previous quenched
(Morningstar and Peardon \cite{Morningstar:1999rf}) 
and dynamical (UKQCD on MILC Asqtad \cite{Steve:2005thes} and 
UKQCD Clover \cite{Hart:2001fp}) determinations with the continuum 
extrapolation performed as in \cite{Morningstar:1999rf}.}
\label{fig:scalargballvsa2}
\end{center}
\end{figure}

In Fig.~\ref{fig:scalargballvsa2} we present our results along with
those from a recent UKQCD study of scalar glueballs 
using staggered fermions 
(MILC configurations)~\cite{Steve:2005thes}, 
a UKQCD study using $\ord{a}$ non-perturbatively improved Wilson
fermions ($N_f=2$)~\cite{Hart:2001fp} 
and the continuum limit result from a quenched anisotropic study of 
the glueball spectrum \cite{Morningstar:1999rf}. 
We perform a simple continuum extrapolation using the form
\BE
\label{eq:gbcontextrap}
r_{0}m(a)=r_0 m_G + b(r_0/a)^4
\EE
where $r_{0}m_G$ and $b$ are the parameters to be determined. 
In~\cite{Morningstar:1999rf}, this was found to work well for all 
glueball states except for the $A_1^{++}$ where
this was thought to be due to strong lattice spacing
dependence caused by proximity to the non-physical phase 
transition in the Wilson fundamental-adjoint plane for
the anisotropic lattice action used.
However,
the alternative form employed in
\cite{Morningstar:1999rf} has four free parameters which 
we are unable to use with just two data points. 
As noted above, it is  ambitious to attempt any kind of continuum
extrapolation using  two data points only 
and we emphasise that the fit with~(\ref{eq:gbcontextrap}) 
has been performed as a guide only. Nevertheless, we that note
good consistency between our continuum value for $r_0m_G$ so obtained,
 $4.32(6)$, and the quenched value
$r_0m_G=4.21(11)(4)$~\cite{Morningstar:1999rf}.
Our extrapolated value corresponds 
to a physical value of $1.83$ GeV approximately .

Our masses seem to show rather weak dependence on $a$ --- certainly
 weaker than that observed in \cite{Hart:2001fp} and of a 
similar strength to that observed in \cite{Morningstar:1999rf}.

\begin{figure}[ht]
\begin{center}
\includegraphics[width=0.9\textwidth,clip]{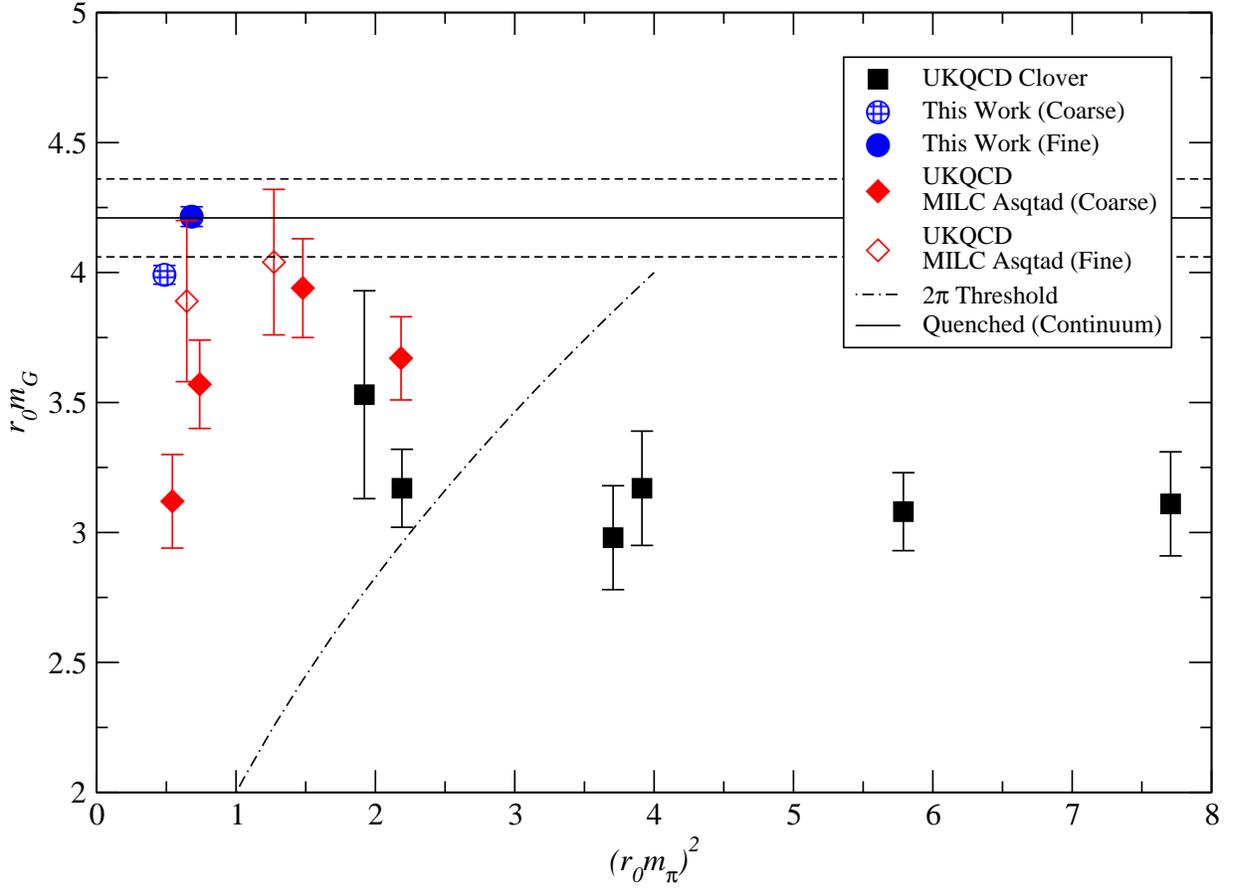}
\caption[Unquenching Effects for Scalar Glueball Measurements]
{Our  measurements of the scalar glueball mass shown with previous  
dynamical (UKQCD on MILC Asqtad~\cite{Steve:2005thes} and UKQCD~Clover \cite{Hart:2001fp})
determinations. 
The $\pi\pi$ threshold is shown (dash-dotted line) for convenience.}
\label{fig:unquenchingscalargb}
\end{center}
\end{figure}

In Fig.~\ref{fig:unquenchingscalargb} we present our results plotted
against the pion mass, shown with the same comparisons from the 
literature as in Fig.~\ref{fig:scalargballvsa2}. 
One might tentatively claim that the glueball mass
increases as one decreases the pion mass, hinting at some underlying mixing
dynamics. However the UKQCD measurements on the coarse MILC 
Asqtad ensembles are rather
spread out and if, as is suspected, the $\ord{a}$ 
improved Wilson measurements are
suppressed by the phase structure of the action used~\cite{Hart:2001fp} 
then they should probably be
discounted and there remains very little trend to study.

\subsection{Tensor and pseudoscalar glueball results}
\label{sse:tens_ps}
\begin{figure}[ht]
\begin{center}
\includegraphics[width=0.75\textwidth,clip]{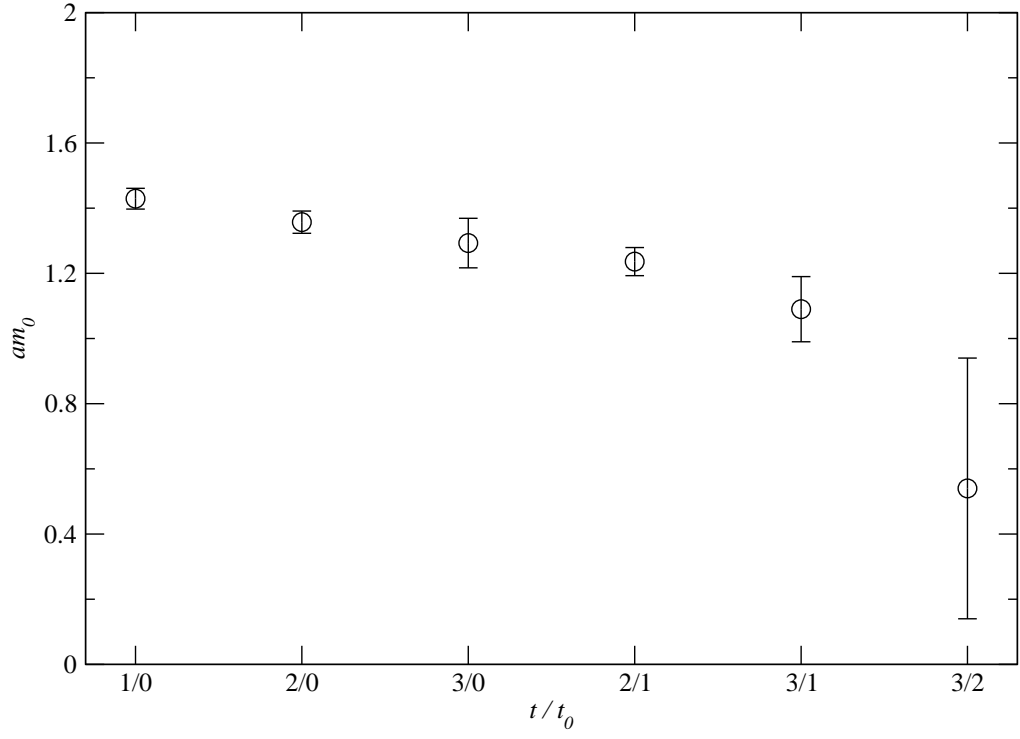}
\caption[Masses from the Variational Eigenvalues: Coarse Pseudoscalar
Operators ($3\times 3$ basis of $\{\op_0^{\Amp},$ $\op_1^{\Amp},$
$\op_1^{\Amp}\}$)]
{The masses extracted from the groundstate variational eigenvalues using
different $t/t_0$ projections performed on a $3\times 3$ matrix of 
correlators formed from the basis of momentum zero 
pseudoscalar glueball operators
$\{\Hop_0^{\Amp},$ $\Hop_1^{\Amp},$ $\Hop_2^{\Amp}\}$ on
the fine ensemble.}
\label{fig:fine_ps_eigenmasses_3x3}
\end{center}
\end{figure}

\begin{figure}[ht]
\begin{center}
\includegraphics[width=0.8\textwidth,clip]
{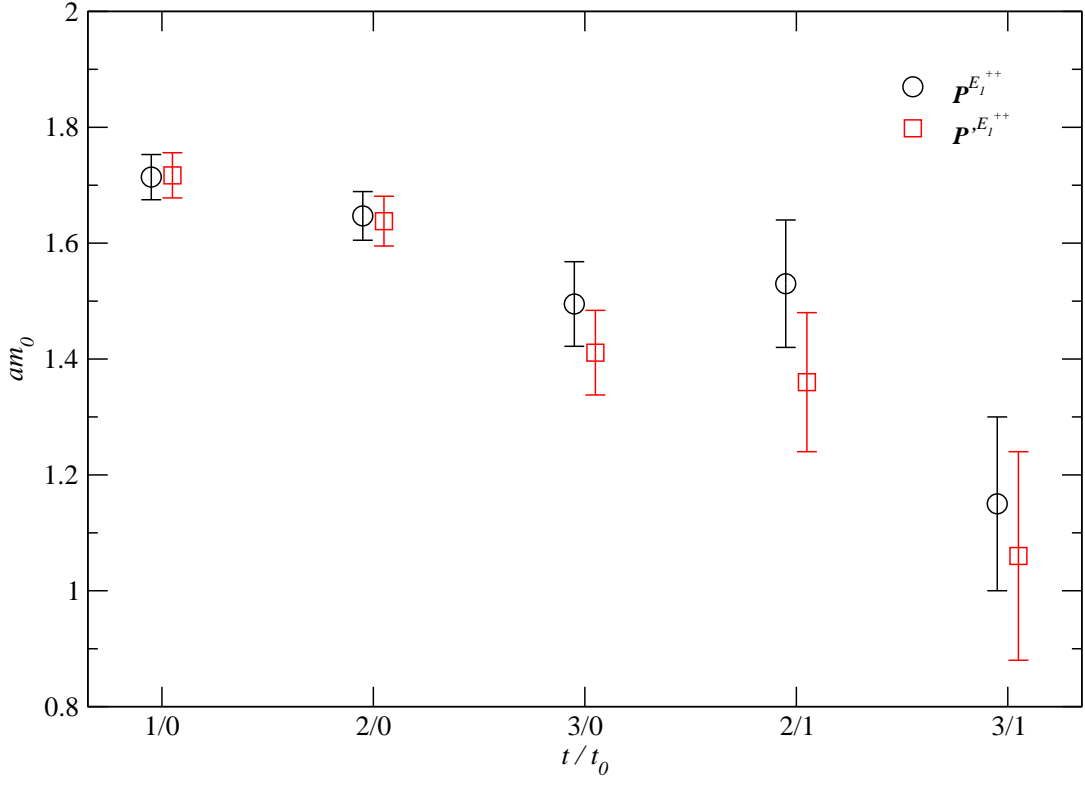}
\caption[Masses from the Variational Eigenvalues: Coarse Tensor
Operators ]
{The masses extracted from the groundstate
 variational eigenvalues for 
different $t/t_0$ projections performed 
on $4\times 4$ matrices of correlators formed using bases of
momentum zero tensor glueball  
operators $\Pop^{\Epp}$ (circles) and $\Pop^{\prime\Epp}$ (squares)
for blocking levels $0$, $1$, $2$ and $3$ in each case (coarse ensemble).}
\label{fig:coarse_tensor_eigmasses_4x4}
\end{center}
\end{figure}

Similar procedures were used to obtain estimates of the ground state and
first excited state in the pseudoscalar and tensor glueball channels. 
For the former we used blocked operators (\ref{eq:psgbop}) 
based on the $\Hop$ operators while for the
latter we used those based on
$\Pop$ -- see (\ref{eq:tensorgbop}) and (\ref{eq:tensorgbop2}).
Sample eigenvalue masses are shown in 
Fig.~\ref{fig:fine_ps_eigenmasses_3x3} and 
Fig.~\ref{fig:coarse_tensor_eigmasses_4x4}
for the pseudoscalar and tensor states respectively.

The global averages are given in Table~\ref{tab:ps_tens_r0}.
\begin{table}[ht]
\centering
\begin{tabular}{|r|c|c|c|c|c|c}
\hline
Result & $am(\Amp)$ & $r_{0}m(\Amp)$ & $am(\Epp)$ & $r_{0}m(\Epp)$ \\
\hline
Coarse -- Ground & 1.560(67) & 5.95(12) & 1.510(13) & 5.756(61) \\
Fine -- Ground & 1.265(17) & 6.399(99) & -- & --\\
Coarse -- Excited & 1.956(65) & 7.46(26) & 1.98(26) & 7.55$\pm$1.01\\
Fine -- Excited & 1.984(77) & 10.04(41)  & -- & --\\
\hline
\end{tabular}
\caption[Ps and tensor Glueball Masses in Units of $r_0$]
{Pseudoscalar and tensor glueball masses 
(ground and first-excited states) from the coarse and fine
ensembles converted into units of the Sommer parameter $r_0$.
Pseudoscalar measurements were made on $3506$/$1998$ configurations of the
coarse/fine ensembles respectively. 
Tensor measurements were made on $2627$ configurations of the
coarse ensemble only.}
\label{tab:ps_tens_r0}
\end{table}
For technical reasons, we were unable to complete measurements
of the tensor state on the fine ensemble.  
For comparison, the quenched continuum limit estimates of $r_0m$
from~\cite{Morningstar:1999rf} are $6.33(13)$ and $5.85(8)$ for pseudoscalar
and tensor glueballs respectively.
So, just as for the scalar channel, there is little evidence of strong 
unquenching effects.

We have conducted a separate study of the $\eta/\eta'$ system using
connected and disconnected meson operators which will be presented
elsewhere. Just for orientation, we note that the continuum
(experimental) values
of $r_0m$ would be $1.32$ ($2.27$) for the $\eta$ ($\eta'$)
respectively.  It is clear that there should be little mixing between
the pseudoscalar glueball and the $\eta'$. However, further excited
states in this singlet channel may mix.

There have been claims that the pseudoscalar 
$\eta(1405)$ meson~\cite{Cheng:2008ss,Gabadadze:1997zc}
contains significant mixtures of the pseudoscalar glueball.
This requires that there is a significant shift of
the mass of the pseudoscalar glueball between quenched and
unquenched QCD, which we don't find in this calculation.
However, note the qualifications accompanying our conclusions below.

\section{Discussion and conclusions}  \label{se:conclusions}
\begin{figure}[ht]
\begin{center}
\includegraphics[width=0.9\textwidth,clip]{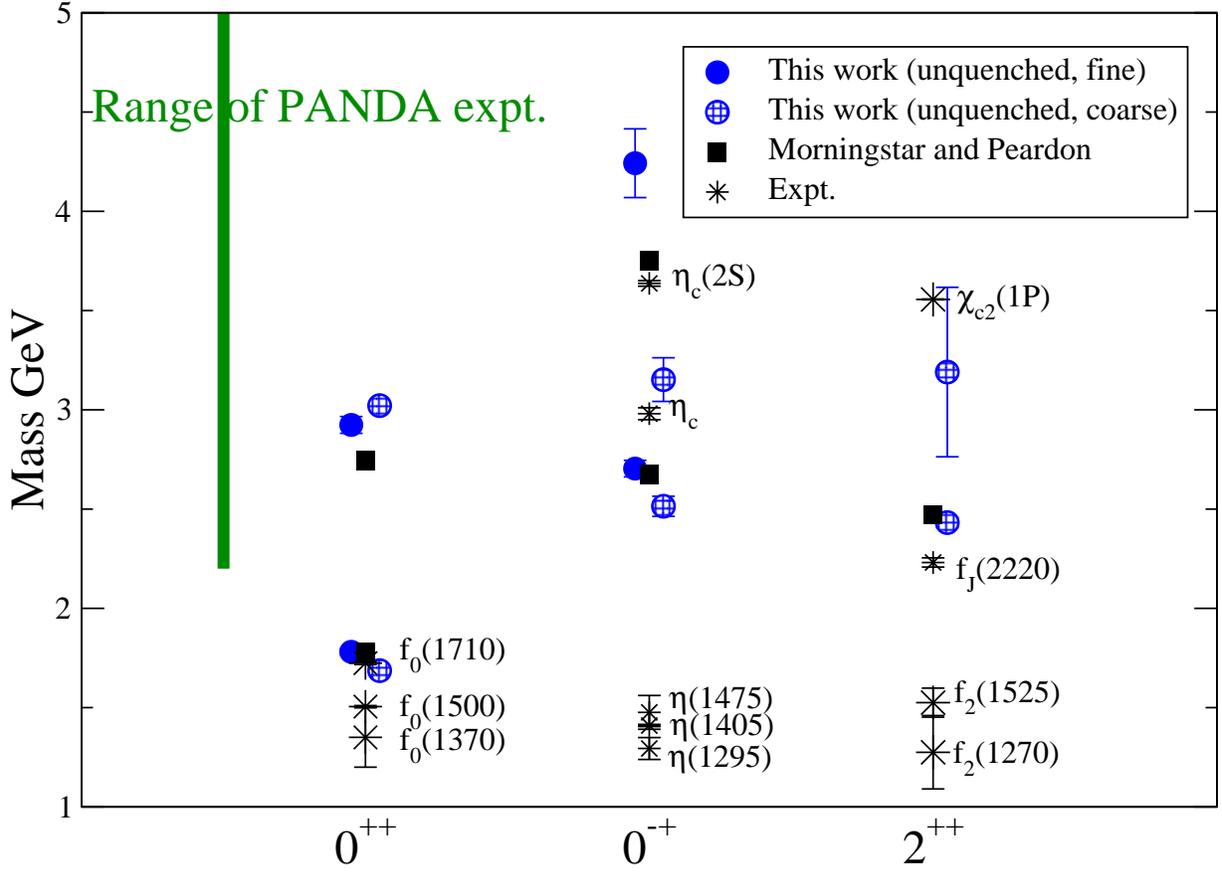}
\caption[Glueball summary]
{Comparison of our unquenched glue-sector predictions
with experimental states listed in the Particle Data Group
listings~\cite{Amsler:2008zzb} and with the quenched continuum limit
predictions of Morningstar and Peardon~\cite{Morningstar:1999rf}. 
The mass range expected to be investigated in
a forthcoming experiment~\cite{Lutz:2009ff}
is also indicated}
\label{fig:Craigplot}
\end{center}
\end{figure}

We have presented the first high statistics study 
of the low-lying glueball sector in $2+1$ flavour QCD. 
We analysed some $5000$ 
configurations of a $24^3\times 64$ lattice with spacing $0.12$ fm and 
$3000$ configurations of a $32^3\times 64$ lattice with spacing $0.09$
fm using improved staggered (Asqtad) dynamical fermions. 
In contrast with earlier glueball analyses using improved Wilson 
fermions, we find no evidence of strong unquenching effects. 
For the $0^{++}$, $0^{-+}$ and $2^{++}$ states we obtain 
mass estimates quite close to the continuum limit masses obtained in
the quenched approximation~\cite{Morningstar:1999rf}.
In Fig.~\ref{fig:Craigplot} 
we show a compilation of $0^{++}$, $0^{-+}$ and $2^{++}$ glueball
candidate states~\cite{Amsler:2008zzb} together with quenched
continuum limit predictions~\cite{Morningstar:1999rf} and our 
unquenched results at fixed lattice spacing and fixed quark mass.

Despite the larger statistics available to us, we still had difficulty
obtaining unambiguous estimates of even the low-lying states.
This was due to at least two main factors - the inherently noisy
correlators and the presence of open decay channels, particularly 
$\pi\pi$. 
In order to reduce the noise, we used a variety of smearing
and blocking techniques giving access to a range of basis states.
We also used a number of different methods based on variational 
techniques: effective masses, transfer matrix eigenvalues and 
multi-channel factorising fits. The effects of the open decay channels 
were exposed at larger euclidean time via the choice of fit ranges and
also via direct evaluation of a subset of mixing matrix elements.

Our overall conclusions are:
\begin{itemize}
\item 
there is little evidence of large unquenching effects on the predicted
low-lying glueball spectrum;
\item
accurate determination of masses requires even larger 
numbers of configurations in comparison
with quenched glueball analyses;
\item
a fuller account must be taken of two meson state contributions
at large euclidean time;
\item
we encountered no problems that could be identified as
resulting from the use of improved staggered fermions -- in particular,
the lattice spacing dependence was weak.
\end{itemize}

\section*{ACKNOWLEDGMENTS}
We are grateful to Chris Michael for advice on fitting techniques and
discussions on mixing.
We acknowledge the use of the SCIDAC-funded Chroma package 
in this analysis~\cite{Edwards:2004sx}.

\bibliographystyle{h-physrev2}
\bibliography{glue}

\end{document}